\documentclass[longauth]{aa}
\usepackage{graphicx}
\usepackage{txfonts}
\usepackage{natbib}
\usepackage{hyperref}
\usepackage{multirow}

\makeatletter
\renewcommand*\aa@pageof{, page \thepage{} of \pageref*{LastPage}}

\newcommand{\xspec}{{\sc xspec}\xspace}

\newcommand{\IXPE}{\textit{IXPE}\xspace}

\newcommand{\INTEGRAL}{\textit{INTEGRAL}\xspace}
\newcommand{\IBIS}{{IBIS/ISGRI}\xspace}
\newcommand{\ATCA}{{ATCA}\xspace}
\newcommand{\VISIR}{{VISIR}\xspace}

\newcommand{\Nustar}{\textit{NuSTAR}\xspace}
\newcommand{\NICER}{{NICER}\xspace}
\newcommand{\REM}{{REM}\xspace}
\newcommand{\LCO}{{LCO}\xspace}
\newcommand{\gx}{{GX~5$-$1}\xspace}

\usepackage{xcolor}

\begin{document}

\title{Discovery of a variable energy-dependent X-ray polarization in the accreting neutron star \gx}

\titlerunning{Discovery of a variable energy-dependent X-ray polarization in the accreting neutron star \gx}
\authorrunning{Fabiani S. et al.} 

\author{
Sergio~Fabiani \inst{\ref{in:INAF-IAPS}} 
\and Fiamma~Capitanio \inst{\ref{in:INAF-IAPS}} 
\and Rosario~Iaria \inst{\ref{in:uni-Palermo}}
\and Juri~Poutanen \inst{\ref{in:UTU}}
\and Andrea~Gnarini \inst{\ref{in:UniRoma3}} 
\and Francesco~Ursini \inst{\ref{in:UniRoma3}}
\and Ruben~Farinelli \inst{\ref{in:INAF-OAS}}
\and Anna~Bobrikova \inst{\ref{in:UTU}} 
\and James~F.~Steiner \inst{\ref{in:CfA}}
\and Jiri~Svoboda \inst{\ref{in:CAS-ASU}}
\and Alessio~Anitra \inst{\ref{in:uni-Palermo}} 
\and Maria~C.~Baglio \inst{\ref{in:AbuDhabi}, \ref{in:INAF-OAB}}  
\and Francesco~Carotenuto \inst{\ref{in:Oxford}}  
\and Melania~Del~Santo \inst{\ref{in:INAF-Palermo}} 
\and Carlo~Ferrigno \inst{\ref{in:Geneva}, \ref{in:INAF-OAB}}
\and Fraser~Lewis \inst{\ref{in:Faulkes},\ref{in:Liverpool}}
\and David~M.~Russell \inst{\ref{in:AbuDhabi}} 
\and Thomas~D.~Russell \inst{\ref{in:INAF-Palermo}} 
\and Jakob~van~den~Eijnden \inst{\ref{in:Warwick}}   
\and Massimo~Cocchi \inst{\ref{in:INAF-OAC}}
\and Alessandro~Di~Marco \inst{\ref{in:INAF-IAPS}}
\and Fabio~La~Monaca \inst{\ref{in:INAF-IAPS}} 
\and Kuan~Liu \inst{\ref{in:GSU}}
\and John~Rankin \inst{\ref{in:INAF-IAPS}} 
\and Martin~C.~Weisskopf \inst{\ref{in:NASA-MSFC}} 
\and Fei~Xie \inst{\ref{in:GSU},\ref{in:INAF-IAPS}} 
\and Stefano~Bianchi \inst{\ref{in:UniRoma3}} 
\and Luciano~Burderi \inst{\ref{in:uni-Monserrato},\ref{in:INFN-Monserrato},\ref{in:INAF-OAC}}
\and Tiziana~Di~Salvo \inst{\ref{in:uni-Palermo},\ref{in:INFN-Monserrato},\ref{in:INAF-OAC}}
\and Elise~Egron \inst{\ref{in:INAF-OAC}}
\and Giulia~Illiano \inst{\ref{in:INAF-OAR}}
\and Philip~Kaaret \inst{\ref{in:NASA-MSFC}}  
\and Giorgio~Matt  \inst{\ref{in:UniRoma3}} 
\and Romana~Miku\v{s}incov\'a \inst{\ref{in:UniRoma3}} 
\and Fabio~Muleri \inst{\ref{in:INAF-IAPS}}  
\and Alessandro~Papitto \inst{\ref{in:INAF-OAR}}
\and Iv\'an~Agudo \inst{\ref{in:CSIC-IAA}}
\and Lucio~A.~Antonelli \inst{\ref{in:INAF-OAR},\ref{in:ASI-SSDC}} 
\and Matteo~Bachetti \inst{\ref{in:INAF-OAC}} 
\and Luca~Baldini  \inst{\ref{in:INFN-PI},      \ref{in:UniPI}} 
\and Wayne~H.~Baumgartner  \inst{\ref{in:NASA-MSFC}} 
\and Ronaldo~Bellazzini  \inst{\ref{in:INFN-PI}} 
\and Stephen~D.~Bongiorno \inst{\ref{in:NASA-MSFC}} 
\and Raffaella~Bonino  \inst{\ref{in:INFN-TO},\ref{in:UniTO}}
\and Alessandro~Brez  \inst{\ref{in:INFN-PI}} 
\and Niccol\`{o}~Bucciantini 
\inst{\ref{in:INAF-Arcetri},\ref{in:UniFI},\ref{in:INFN-FI}} 
\and Simone~Castellano \inst{\ref{in:INFN-PI}}  
\and Elisabetta~Cavazzuti \inst{\ref{in:ASI}} 
\and Chien-Ting~Chen \inst{\ref{in:USRA-MSFC}}
\and Stefano~Ciprini \inst{\ref{in:INFN-Roma2},\ref{in:ASI-SSDC}}
\and Enrico~Costa \inst{\ref{in:INAF-IAPS}} 
\and Alessandra~De~Rosa \inst{\ref{in:INAF-IAPS}} 
\and Ettore~Del~Monte \inst{\ref{in:INAF-IAPS}} 
\and Laura~Di~Gesu \inst{\ref{in:ASI}} 
\and Niccol\`{o}~Di~Lalla \inst{\ref{in:Stanford}}
\and Immacolata~Donnarumma \inst{\ref{in:ASI}}
\and Victor~Doroshenko \inst{\ref{in:Tub}}
\and Michal~Dov\v{c}iak \inst{\ref{in:CAS-ASU}}
\and Steven~R.~Ehlert \inst{\ref{in:NASA-MSFC}}  
\and Teruaki~Enoto \inst{\ref{in:RIKEN}}
\and Yuri~Evangelista \inst{\ref{in:INAF-IAPS}}
\and Riccardo~Ferrazzoli \inst{\ref{in:INAF-IAPS}} 
\and Javier~A.~Garcia \inst{\ref{in:Caltech}}
\and Shuichi~Gunji \inst{\ref{in:Yamagata}} 
\and Kiyoshi~Hayashida \inst{\ref{in:Osaka}} \thanks{Decessed.}
\and Jeremy~Heyl \inst{\ref{in:UBC}}
\and Wataru~Iwakiri \inst{\ref{in:Chiba}} 
\and Svetlana~G.~Jorstad \inst{\ref{in:BU},\ref{in:SPBU}} 
\and Vladimir~Karas \inst{\ref{in:CAS-ASU}}
\and Fabian~Kislat \inst{\ref{in:UNH}} 
\and Takao~Kitaguchi  \inst{\ref{in:RIKEN}} 
\and Jeffery~J.~Kolodziejczak \inst{\ref{in:NASA-MSFC}} 
\and Henric~Krawczynski  \inst{\ref{in:WUStL}}
\and Luca~Latronico  \inst{\ref{in:INFN-TO}} 
\and Ioannis~Liodakis \inst{\ref{in:FINCA}}
\and Simone~Maldera \inst{\ref{in:INFN-TO}}  
\and Alberto~Manfreda \inst{\ref{INFN-NA}}
\and Fr\'{e}d\'{e}ric~Marin \inst{\ref{in:Strasbourg}} 
\and Andrea~Marinucci \inst{\ref{in:ASI}} 
\and Alan~P.~Marscher \inst{\ref{in:BU}} 
\and Herman~L.~Marshall \inst{\ref{in:MIT}}
\and Francesco~Massaro \inst{\ref{in:INFN-TO},\ref{in:UniTO}}  
\and Ikuyuki~Mitsuishi \inst{\ref{in:Nagoya}} 
\and Tsunefumi~Mizuno \inst{\ref{in:Hiroshima}} 
\and Michela~Negro \inst{\ref{in:UMBC},\ref{in:NASA-GSFC},\ref{in:CRESST}} 
\and Chi-Yung~Ng \inst{\ref{in:HKU}}
\and Stephen~L.~O'Dell \inst{\ref{in:NASA-MSFC}}  
\and Nicola~Omodei \inst{\ref{in:Stanford}}
\and Chiara~Oppedisano \inst{\ref{in:INFN-TO}}  
\and George~G.~Pavlov \inst{\ref{in:PSU}}
\and Abel~L.~Peirson \inst{\ref{in:Stanford}}
\and Matteo~Perri \inst{\ref{in:ASI-SSDC},\ref{in:INAF-OAR}}
\and Melissa~Pesce-Rollins \inst{\ref{in:INFN-PI}} 
\and Pierre-Olivier~Petrucci \inst{\ref{in:Grenoble}} 
\and Maura~Pilia \inst{\ref{in:INAF-OAC}} 
\and Andrea~Possenti \inst{\ref{in:INAF-OAC}} 
\and Simonetta~Puccetti \inst{\ref{in:ASI-SSDC}}
\and Brian~D.~Ramsey \inst{\ref{in:NASA-MSFC}}  
\and Ajay~Ratheesh \inst{\ref{in:INAF-IAPS}} 
\and Oliver~J.~Roberts \inst{\ref{in:USRA-MSFC}}
\and Roger~W.~Romani \inst{\ref{in:Stanford}}
\and Carmelo~Sgr\`o \inst{\ref{in:INFN-PI}}  
\and Patrick~Slane \inst{\ref{in:CfA}}  
\and Paolo~Soffitta \inst{\ref{in:INAF-IAPS}} 
\and Gloria~Spandre \inst{\ref{in:INFN-PI}} 
\and Douglas~A.~Swartz \inst{\ref{in:USRA-MSFC}}
\and Toru~Tamagawa \inst{\ref{in:RIKEN}}
\and Fabrizio~Tavecchio \inst{\ref{in:INAF-OAB}}
\and Roberto~Taverna \inst{\ref{in:UniPD}} 
\and Yuzuru~Tawara \inst{\ref{in:Nagoya}}
\and Allyn~F.~Tennant \inst{\ref{in:NASA-MSFC}}  
\and Nicholas~E.~Thomas \inst{\ref{in:NASA-MSFC}}  
\and Francesco~Tombesi  \inst{\ref{in:UniRoma2},\ref{in:INFN-Roma2},\ref{in:UMd}}
\and Alessio~Trois \inst{\ref{in:INAF-OAC}}
\and Sergey~S.~Tsygankov \inst{\ref{in:UTU}}
\and Roberto~Turolla \inst{\ref{in:UniPD},\ref{in:MSSL}}
\and Jacco~Vink \inst{\ref{in:Amsterdam}}
\and Kinwah~Wu \inst{\ref{in:MSSL}}
\and Silvia~Zane  \inst{\ref{in:MSSL}}
}

\institute{
INAF Istituto di Astrofisica e Planetologia Spaziali, Via del Fosso del Cavaliere 100, 00133 Roma, Italy \label{in:INAF-IAPS}
\email{sergio.fabiani@inaf.it}
\and Universit\`a degli Studi di Palermo, Dipartimento di Fisica e Chimica, Emilio Segr\`e, via Archirafi 36 I-90123 Palermo, Italy \label{in:uni-Palermo}
\and Department of Physics and Astronomy, FI-20014 University of Turku,  Finland \label{in:UTU} \\ 
\and 
Dipartimento di Matematica e Fisica, Universit\`a degli Studi Roma Tre, via della Vasca Navale 84, 00146 Roma, Italy  \label{in:UniRoma3}
\and
INAF Osservatorio di Astrofisica e Scienza dello Spazio di Bologna, Via P. Gobetti 101, I-40129 Bologna, Italy \label{in:INAF-OAS} 
\and 
Center for Astrophysics, Harvard \& Smithsonian, 60 Garden St, Cambridge, MA 02138, USA \label{in:CfA} 
\and 
Astronomical Institute of the Czech Academy of Sciences, Bo\v{c}n\'{i} II 1401/1, 14100 Praha 4, Czech Republic \label{in:CAS-ASU}
\and Center for Astro, Particle and Planetary Physics, New York University Abu Dhabi, PO Box 129188, Abu Dhabi, UAE \label{in:AbuDhabi} 
\and INAF Osservatorio Astronomico di Brera, via E. Bianchi 46, 23807 Merate (LC), Italy \label{in:INAF-OAB}
\and  Department of Physics, University of Warwick, Coventry CV4 7AL, UK \label{in:Warwick} 
\and Astrophysics, Department of Physics, University of Oxford, Keble Road, Oxford OX1 3RH \label{in:Oxford}
\and 
INAF Istituto di Astrofisica Spaziale e Fisica Cosmica, Via U. La Malfa 153, I-90146 Palermo, Italy \label{in:INAF-Palermo} 
\and Department of Astronomy, University of Geneva, Ch. d'Ecogia 16, CH-1290 Versoix, Geneva, Switzerland \label{in:Geneva}
\and Faulkes Telescope Project, School of Physics and Astronomy, Cardiff University, The Parade, Cardiff CF24 3AA, Wales, UK \label{in:Faulkes} 
\and Astrophysics Research Institute, Liverpool John Moores University, 146 Brownlow Hill, Liverpool L3 5RF, UK \label{in:Liverpool}
\and
INAF Osservatorio Astronomico di Cagliari, Via della Scienza 5, 09047 Selargius (CA), Italy  \label{in:INAF-OAC}
\and 
Guangxi Key Laboratory for Relativistic Astrophysics, School of Physical Science and Technology, Guangxi University, Nanning 530004, China \label{in:GSU}
\and 
NASA Marshall Space Flight Center, Huntsville, AL 35812, USA \label{in:NASA-MSFC}
\and  Dipartimento di Fisica, Universit\`a degli Studi di Cagliari, SP Monserrato-Sestu, KM 0.7, Monserrato, I-09042 Italy \label{in:uni-Monserrato}
\and INFN, Sezione di Cagliari, Cittadella Universitaria, I-09042 Monserrato, CA, Italy \label{in:INFN-Monserrato}
\and 
INAF Osservatorio Astronomico di Roma, Via Frascati 33, 00040 Monte Porzio Catone (RM), Italy \label{in:INAF-OAR} 
\and 
Instituto de Astrof\'{i}sicade Andaluc\'{i}a -- CSIC, Glorieta de la Astronom\'{i}a s/n, 18008 Granada, Spain \label{in:CSIC-IAA}
\and 
Space Science Data Center, Agenzia Spaziale Italiana, Via del Politecnico snc, 00133 Roma, Italy \label{in:ASI-SSDC}
\and 
Istituto Nazionale di Fisica Nucleare, Sezione di Pisa, Largo B. Pontecorvo 3, 56127 Pisa, Italy \label{in:INFN-PI}
\and  
Dipartimento di Fisica, Universit\`{a} di Pisa, Largo B. Pontecorvo 3, 56127 Pisa, Italy \label{in:UniPI} 
\and  
Istituto Nazionale di Fisica Nucleare, Sezione di Torino, Via Pietro Giuria 1, 10125 Torino, Italy  \label{in:INFN-TO}      
\and  
Dipartimento di Fisica, Universit\`{a} degli Studi di Torino, Via Pietro Giuria 1, 10125 Torino, Italy \label{in:UniTO} 
\and   
INAF Osservatorio Astrofisico di Arcetri, Largo Enrico Fermi 5, 50125 Firenze, Italy 
\label{in:INAF-Arcetri} 
\and  
Dipartimento di Fisica e Astronomia, Universit\`{a} degli Studi di Firenze, Via Sansone 1, 50019 Sesto Fiorentino (FI), Italy \label{in:UniFI} 
\and   
Istituto Nazionale di Fisica Nucleare, Sezione di Firenze, Via Sansone 1, 50019 Sesto Fiorentino (FI), Italy \label{in:INFN-FI}
\and 
Agenzia Spaziale Italiana, Via del Politecnico snc, 00133 Roma, Italy \label{in:ASI}
\and 
Science and Technology Institute, Universities Space Research Association, Huntsville, AL 35805, USA \label{in:USRA-MSFC}
\and 
Istituto Nazionale di Fisica Nucleare, Sezione di Roma ``Tor Vergata'', Via della Ricerca Scientifica 1, 00133 Roma, Italy 
\label{in:INFN-Roma2}
\and 
Department of Physics and Kavli Institute for Particle Astrophysics and Cosmology, Stanford University, Stanford, California 94305, USA  \label{in:Stanford}
\and
Institut f\"ur Astronomie und Astrophysik, Universit\"at T\"ubingen, Sand 1, D-72076 T\"ubingen, Germany \label{in:Tub}
\and 
RIKEN Cluster for Pioneering Research, 2-1 Hirosawa, Wako, Saitama 351-0198, Japan \label{in:RIKEN}
\and 
California Institute of Technology, Pasadena, CA 91125, USA \label{in:Caltech}
\and 
Yamagata University,1-4-12 Kojirakawa-machi, Yamagata-shi 990-8560, Japan \label{in:Yamagata}
\and 
Osaka University, 1-1 Yamadaoka, Suita, Osaka 565-0871, Japan \label{in:Osaka}
\and 
University of British Columbia, Vancouver, BC V6T 1Z4, Canada \label{in:UBC}
\and 
International Center for Hadron Astrophysics, Chiba University, Chiba 263-8522, Japan \label{in:Chiba}
\and
Institute for Astrophysical Research, Boston University, 725 Commonwealth Avenue, Boston, MA 02215, USA \label{in:BU} 
\and 
Department of Astrophysics, St. Petersburg State University, Universitetsky pr. 28, Petrodvoretz, 198504 St. Petersburg, Russia \label{in:SPBU} 
\and 
Department of Physics and Astronomy and Space Science Center, University of New Hampshire, Durham, NH 03824, USA \label{in:UNH} 
\and 
Physics Department and McDonnell Center for the Space Sciences, Washington University in St. Louis, St. Louis, MO 63130, USA \label{in:WUStL}
\and 
Finnish Centre for Astronomy with ESO,  20014 University of Turku, Finland \label{in:FINCA}
\and 
Istituto Nazionale di Fisica Nucleare, Sezione di Napoli, Strada Comunale Cinthia, 80126 Napoli, Italy \label{INFN-NA}
\and 
Universit\'{e} de Strasbourg, CNRS, Observatoire Astronomique de Strasbourg, UMR 7550, 67000 Strasbourg, France \label{in:Strasbourg}
\and 
MIT Kavli Institute for Astrophysics and Space Research, Massachusetts Institute of Technology, 77 Massachusetts Avenue, Cambridge, MA 02139, USA \label{in:MIT}
\and 
Graduate School of Science, Division of Particle and Astrophysical Science, Nagoya University, Furo-cho, Chikusa-ku, Nagoya, Aichi 464-8602, Japan \label{in:Nagoya}
\and 
Hiroshima Astrophysical Science Center, Hiroshima University, 1-3-1 Kagamiyama, Higashi-Hiroshima, Hiroshima 739-8526, Japan \label{in:Hiroshima}
\and
University of Maryland, Baltimore County, Baltimore, MD 21250, USA \label{in:UMBC}
\and 
NASA Goddard Space Flight Center, Greenbelt, MD 20771, USA  \label{in:NASA-GSFC}
\and 
Center for Research and Exploration in Space Science and Technology, NASA/GSFC, Greenbelt, MD 20771, USA  \label{in:CRESST}
\and 
Department of Physics, University of Hong Kong, Pokfulam, Hong Kong \label{in:HKU}
\and 
Department of Astronomy and Astrophysics, Pennsylvania State University, University Park, PA 16801, USA \label{in:PSU}
\and 
Universit\'{e} Grenoble Alpes, CNRS, IPAG, 38000 Grenoble, France \label{in:Grenoble}
\and 
Dipartimento di Fisica e Astronomia, Universit\`{a} degli Studi di Padova, Via Marzolo 8, 35131 Padova, Italy \label{in:UniPD}
\and
Dipartimento di Fisica, Universit\`{a} degli Studi di Roma ``Tor Vergata'', Via della Ricerca Scientifica 1, 00133 Roma, Italy \label{in:UniRoma2}
\and
Department of Astronomy, University of Maryland, College Park, Maryland 20742, USA \label{in:UMd}
\and 
Mullard Space Science Laboratory, University College London, Holmbury St Mary, Dorking, Surrey RH5 6NT, UK \label{in:MSSL}
\and 
Anton Pannekoek Institute for Astronomy \& GRAPPA, University of Amsterdam, Science Park 904, 1098 XH Amsterdam, The Netherlands  \label{in:Amsterdam}
}


\abstract
{\\We report on the coordinated observations of the neutron star low-mass X-ray binary (NS-LMXB) \gx in X-rays (\IXPE, \NICER, \Nustar, and \INTEGRAL), optical (REM and LCO), near-infrared (REM), mid-infrared (VLT VISIR), and radio (ATCA).
This Z-source was observed by \IXPE twice in March--April 2023 (Obs. 1 and 2). 
In the radio band the source was detected, but only upper limits to the linear polarization were obtained at a $3\sigma$ level of 6.1\% at 5.5 GHz and 5.9\% at 9 GHz in Obs.~1 and 12.5\% at 5.5~GHz and 20\% at 9~GHz in Obs.~2. The mid-IR, near-IR, and optical observations suggest the presence of a compact jet that peaks in the mid- or far-IR.
The X-ray polarization degree was found to be $3.7\% \pm 0.4 \%$ (at 90\% confidence level) during Obs.~1 when the source was in the horizontal branch of the Z-track and $1.8\% \pm 0.4 \%$ during Obs.~2 when the source was in the normal-flaring branch. These results confirm the variation in polarization degree as a function of the position of the source in the color-color diagram, as for previously observed Z-track sources (Cyg~X-2 and XTE~1701$-$462). Evidence of a variation in the polarization angle of $\sim 20\degr$ with energy is found in both observations, likely related to the different, nonorthogonal polarization angles of the disk and Comptonization components, which peak at different energies.}   


   \maketitle
%

\section{Introduction}
\label{sec:intro}

Persistent neutron star low-mass X-ray binaries (NS-LMXBs) are among the X-ray astronomical objects that the \textit{Imaging X-ray Polarimetry Explorer}~(\IXPE, \citealt{Weisskopf2023,Weisskopf22, Soffitta21}) is investigating.
Four of them were   observed by \IXPE during the  first year of the
campaign, namely the  Z-source Cyg~X-2; the peculiar Z-Atoll transient XTE J1701$-$462; and two bright soft-state Atoll-sources, GS~1826$-$238 and GX~9$+$9. The source classification (Z or Atoll) is based on the tracks that they draw on the color-color diagram \citep[CCD; see, e.g.,][]{hasinger89,vdK95}. GS~1826$-$238 data are compatible with a null polarization with an upper limit on the polarization degree (PD) of 1.3\% \citep{Capitanio23}, while Cyg~X-2 \citep{Farinelli23} and GX~9+9 
\citep{Chatterjee23gx99,Ursini2023} have shown a statistically significant linear polarization with the PD of $\sim$2\% and $\sim$1.5\%, respectively. Strong variations in PD were reported for XTE~J1701$-$462, observed twice, that showed a high PD$\sim4.5$\% in the first observation and a PD compatible with a null polarization in the second   \citep{Cocchi2023}.


\begin{table*} 
\centering
\caption{List of observations.}
\label{tab:obslog}
\begin{tabular}{lccccc}
\hline
\hline           
\text{Telescope} & \text{ObsID} & \text{Obs. Start} & \text{Obs. Stop} & \text{Net Exposure (ks)}& \text{Notes} \\
\hline
\multicolumn{5}{c}{\IXPE Obs. 1} \\
\hline
\text{\IXPE} & 02002799 & \text{2023-03-21, 04:16:14} & \text{2023-03-22, 05:02:52} & 48.6&  obs. segment  1  \\
\text{\NICER} & \text{6010230101/2}& \text{2023-03-21, 03:41:20} & \text{2023-03-22, 04:50:20} & 13.1&  \\
\text{\Nustar} & 90902310002 & \text{2023-03-21, 16:41:48} & \text{2023-03-22, 08:04:34} & 12.6&    \\
\text{\INTEGRAL}& 2070006/0001 & \text{2023-03-21, 03:58:07} & \text{2023-03-22, 04:28:46} & 40.1&    \\
\text{\REM} & -- & \text{2023-03-22, 05:25:24} & \text{2023-03-22, 07:33:05} & --&  see Sect.~\ref{sec:rem} for exp. details   \\
\text{\LCO} & -- & \text{2023-03-22, 16:33:07} &  \text{2023-03-22, 16:38:07} & --&  "  \\
            \hline

\multicolumn{5}{c}{No \IXPE Obs.} \\
            \hline
\text{\NICER} & 6010230103/4& \text{2023-03-24 18:19:20} & \text{2023-03-25 00:44:20} & 5.2&  \\
\text{\ATCA} & -- & \text{2023-03-24, 15:21:20} & \text{2023-03-25, 01:55:50} & --&  see Sect.~\ref{sec:atca} for exp. details\\
\text{\VISIR} &  110.2448 & \text{2023-03-28, 08:31:00 } & \text{2023-03-28, 09:21:00} & --&  see Sect.~\ref{sec:visir} for exp. details \\
\text{\LCO} & -- & \text{2023-03-29, 15:50:11} & \text{2023-03-31, 15:43:07} & --&  see Sect.~\ref{sec:rem} for exp. details  \\
\text{\VISIR} &  110.2448 & \text{2023-03-31, 07:46:00 } & \text{2023-03-31, 08:42:00} & --& see Sect.~\ref{sec:visir} for exp. details  \\

            \hline
\multicolumn{5}{c}{\IXPE Obs. 2} \\
\hline
\text{\IXPE} & 02002799 & \text{2023-04-13, 23:43:42} & \text{ 2023-04-15, 00:37:32 } & 47.1&  obs. segment  2  \\
\text{\NICER} & 6010230105/6 & \text{2023-04-13 16:39:04} & \text{2023-04-15, 23:58:20} & \text{13.1} &  \\
\text{\Nustar} & 90902310004/6  & \text{2023-04-13, 15:57:07} & \text{2023-04-14, 23:23:24} & 15.7&    \\
\text{\INTEGRAL}& multiple & \text{2023-04-13, 03:51:17} & \text{2023-04-15 08:53:00} & 85.0 &    \\
\text{\REM} & -- & \text{2023-04-14, 03:16:55} & \text{2023-04-14, 06:31:56} & --&   see Sect.~\ref{sec:rem} for exp. details   \\
\text{\LCO} & -- & \text{2023-04-14, 05:35:19} & \text{2023-04-14, 23:23:55} & --&  "  \\
\text{\ATCA} & -- & \text{2023-04-14, 11:32:00} & \text{2023-04-14, 17:57:40} & --&  see Sect.~\ref{sec:atca} for exp. details\  \\
\hline
\end{tabular}
\tablefoot{
The three table sections comprise the observations related to \IXPE Obs.~1,  Obs.~2, and observations in between performed at other observatories.}
\end{table*}

\gx is a Galactic Z-source \citep{Kuulkers1994,Jonker2002} located near the Galactic center. It is a radio source with radio emission most likely originating from a compact jet \citep{Fender2000}. The radio counterpart allowed an accurate localization that, despite optical obscuration and the crowded field near the Galactic center, has led to the determination of a likely infrared companion candidate \citep{Jonker2000}.
Until the early 1990s \gx X-ray data were likely contaminated by the black hole LMXB GRS~1758$-$258, located only 40\arcmin\ away.
\citet{Sunyaev1991} and \citet{Gilfanov1993} were able to resolve two sources, showing that \gx was $\sim$30--50  times brighter than GRS~1758$-$258 below 20~keV. \gx has not shown any X-ray pulsations or X-ray bursts \citep{Paizis2005}.

An X-ray halo due to scattering of X-ray photons is clearly revealed in the \textit{Chandra} image  \citep{Smith2006,Clark2018}. Such a halo arises due to the presence of multiple clouds along the line of sight.

In X-rays and $\gamma$-rays, \citet{Paizis2005} studied one year of \INTEGRAL data, which covered the entire Z-track, mainly the horizontal branch (HB) and normal branch (NB).
ISGRI and JEM-X average spectra showed  a clear hard X-ray emission above 20~keV, not detected previously and compatible with thermal Comptonization of soft photons from a hot optically thin plasma in the vicinity of the NS. However, \citet{Paizis2005} were not able to constrain the temperature of the Comptonizing plasma. They assessed the compatibility of the \gx energy spectrum with the  eastern \citep{Mitsuda1984} and western \citep{White1986} models, and found the former to be  physically more meaningful, describing the spectral flattening above $\sim20$~keV as a Comptonized hard-tail emission. 
\citet{Paizis2006} described the \gx energy spectrum in the 20--100~keV energy band observed by \INTEGRAL IBIS/ISGRI with a Comptonization component (\texttt{comptt}; \citealt{Titarchuk1994}) plus a power law to account for the hard X-ray emission. Such a high-energy tail was first detected by \citet{Asai1994}, although a possible contamination from the nearby black hole GRS~1758$-$258 could not be excluded in this latter case.

\citet{Paizis2006} highlighted the presence of a correlation between the X-ray spectral states and the radio emission.
Steady radio emission is associated with a low-hard state \citep[typical for Atoll sources;][]{Fender2000,Migliari2006}.
These sources brighten considerably during intermediate states, often showing bright, transient flares (HB of Z-sources), before quenching during the very soft states in the NB and FB branches of Z-sources \citep{Fender2000,DiSalvo2002}. 
Moreover, \citet{Paizis2006} reported that the radio flux is positively correlated with the flux of the hard tail in the 40--100~keV energy range. They suggested that this correlation is related to the acceleration of electrons along open field lines in the NS magnetosphere at the base of the jet seen in the radio.
\citet{Berendsen2000} reported upper limits to the radio linear polarization of 33\% at 6.3\,cm and 23\% at 3.5\,cm not sufficient to constrain the emission mechanism or the optical depth of the jet.

\begin{figure*}
\centering
\includegraphics[width=0.99\textwidth]{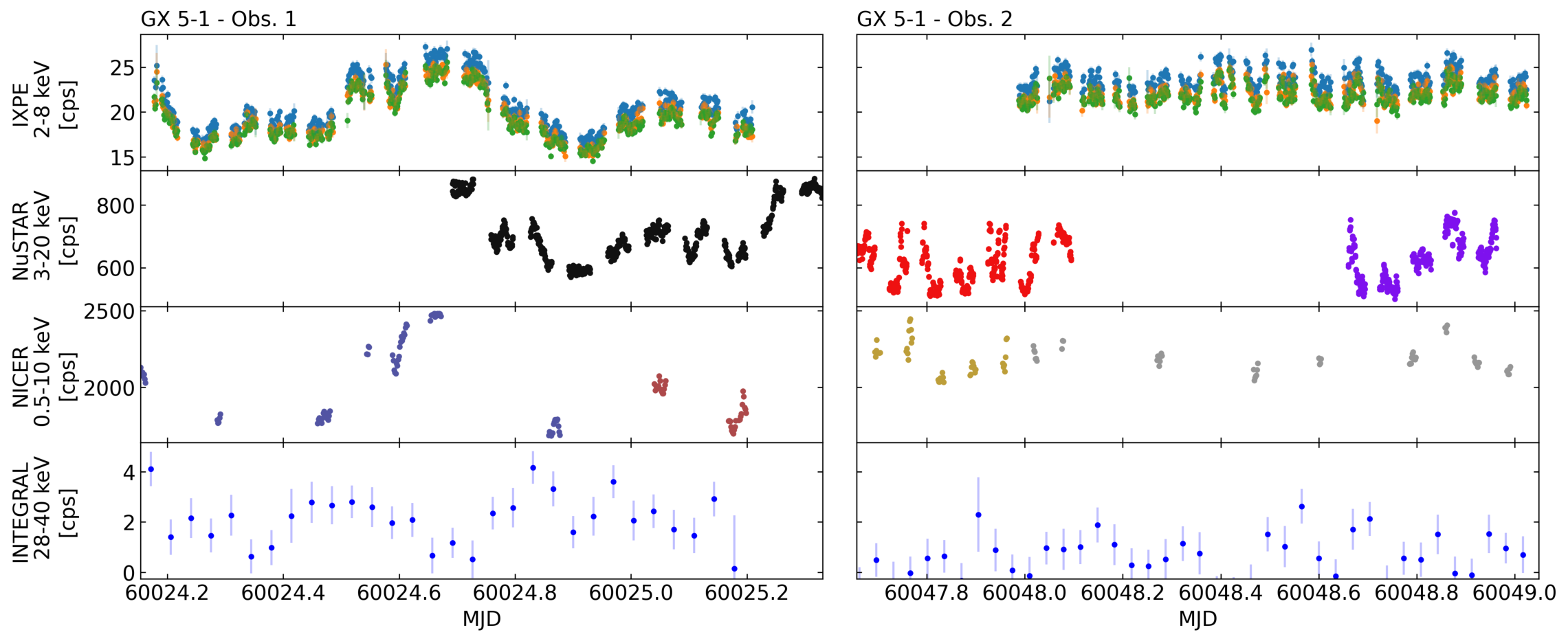}
\caption{Light curves of \Nustar, \NICER, and \IBIS during \IXPE Obs.~1 (left panel) and Obs.~2. (right panel). In the top row of each panel the three \IXPE DU light curves are shown (DU1 blue, DU2 orange, and DU3 green). 
In the left panel the \NICER soft purple and dark red points correspond to ObsID 6010230101 and 6010230102. In the right panel the \Nustar red and purple points correspond to ObsIDs 90902310004 and 90902310006. The \NICER gray points correspond to ObsID 6010230106.}
\label{fig:lc}
\end{figure*}

\section{Observations and data reduction}
\label{sec:datareduc}

\subsection{IXPE}
\label{sec:ixpe}
The X-ray polarimeter \IXPE \citep{Weisskopf2023,Weisskopf22,Soffitta21} observed \gx twice (ObsID 02002799) from March 21 to 22 and from April 13 to 15, 2023 (see Table \ref{tab:obslog} and Fig.~\ref{fig:lc} for the light curves), with a nominal integration time of 50~ks each.
\IXPE provides timing, imaging, spectroscopic, and polarimetric data in the 2--8 keV band.
Data reduction and analysis were performed by means of the \textsc{ixpeobssim} software version 30.5.0 \citep{Baldini22} and the \textsc{heasoft} package version 6.31.1 \citep{2014ascl.soft08004N}.
Data were filtered by using \textsc{ixpeobssim} tools \texttt{xpselect} and binned\footnote{The default energy binning is 40~eV.} with \texttt{xpbin} to produce images and $I$, $Q$, and $U$ energy spectra for spectropolarimetric analysis performed with \textsc{xspec} version 12.13.0c \citep{Arnaud96}.
We used the latest version 12 of the \IXPE response matrices available at the \textsc{ixpeobssim} public repository\footnote{https://github.com/lucabaldini/ixpeobssim} (also available at the HEASARC archive).
A circular source extraction region of 60\arcsec\ in radius was selected from the image for each  of the three detector units (DUs). No background subtraction was applied due to the high count rate of the source ($\sim$20--25 cts~s$^{-1}$ per DU) \citep[see][]{DiMarco23}.
Only \textsc{xspec}  currently allows a weighted analysis \citep{DiMarco2022} in which a weight is assigned to each photo-electron track recorded by the DUs depending on the shape of the charge distribution.

The normalized Stokes parameters $q=Q/I$ and $u=U/I$, the PD, and polarization angle (PA) with their uncertainties can be calculated by using the model-independent \texttt{pcube} binning algorithm of \textsc{ixpeobssim}.
On the other hand, PD and PA obtained with \textsc{xspec} require the definition of a spectropolarimetric model.
Because the PD and PA are not independent, the appropriate way to report the results is by means of contour plots at certain confidence levels for the joint measurements of the two parameters.
We report the  \textsc{xspec} (PD, PA) contour plots  obtained by using the \texttt{steppar} command, while the contour plots associated with the \textsc{ixpeobssim} analysis were obtained from the statistics of the number of counts \citep{Weisskopf2010,Strohmayer2013,Muleri2022}.

The spectropolarimetric analysis was carried out by taking into account the current \IXPE effective area instrument response function (arf), which may not be as accurate as possible at energies above 6 keV where there is a significant roll-off in the spectral response. Because the high-energy part of the \IXPE band is of special interest for our study, we estimate the \IXPE spectral systematic uncertainties to be   3\% in Obs. 1 and 2\% in Obs. 2, based on the comparison with the  \NICER, \Nustar, and \IBIS energy spectra.
The spectral models for both the \IXPE observations are frozen based on the  spectral analysis of these other observatories.
We used the \textsc{xspec} gain fit tool for the \IXPE data only, to shift the energies on which the response matrix is defined and to match the effective area curve during the fit procedure. The spectropolarimetric fit of the \IXPE data was carried out freeing the three DU normalization constants: the gain slope, the gain offset, and the polarimetric parameters.\footnote{The normalization constants, gain slope, and offset parameters are identical between the  $I$, $Q$, and $U$ spectra of the same DU. The PA and PD parameters of those spectra are also identical  for each DU.}

\subsection{\Nustar}
\label{sec:nustar}
\gx was observed by \Nustar \citep{Harrison2013} on March 21 and twice on April 13--14, 2023.
All the relevant observation times, sequence IDs, and exposure times are provided in Table \ref{tab:obslog}, and the corresponding light curves are shown in Fig.~\ref{fig:lc}.
The unfiltered event files were processed with the \Nustar Data Analysis Software (\textsc{nustardas} v.2.1.2) to produce the cleaned and calibrated level 2 data using the latest calibration files (CALDB v.20221130) and the standard filtering criteria with the \texttt{nupipeline} task, where \texttt{statusexpr"STATUS==b0000xxx00xxxx000"} was set, due to the source flux exceeding 100 counts s$^{-1}$.
The spectra and light curves were extracted using the \texttt{nuproducts} task, selecting a circular region of 60\arcsec\ in radius centered on the source.

\subsection{\NICER}
\label{sec:nicer}
\NICER \citep{Gendreau2016} observed \gx between March 21--25 and  April 13--14, 2023.
The observations identified with IDs 6010230101, 6010230102, and 
6010230106 were included in the spectropolarimetric analysis because they are 
simultaneous with the \IXPE observations.
The calibrated and cleaned files were extracted by using the standard \texttt{nicerl2} command of the \NICER Data Analysis Software (\textsc{nicerdas} v.10) together with the latest calibration files (CALDB v.20221001).
The spectra and the light curves were then obtained with the \texttt{nicerl3-spect} and \texttt{nicerl3-lc} tasks, while the background was computed using the SCORPEON\footnote{\url{https://heasarc.gsfc.nasa.gov/docs/nicer/analysis_threads/scorpeon-overview/}} model.
The light curves were obtained including 50 \NICER FPMs available during the ObsIDs included in the analysis (out of 52 available, excluding noisy detectors ID~14 and ID~34). Thus, no discontinuities in the \NICER light curves are present (see Fig. \ref{fig:lc}).

During the contemporary observation with \IXPE there were two significant increases in count rate in the \NICER data, up to twice the typical value. These events were coincident with two solar flares; the first   was a C-class flare peaking at about 60048.5389~MJD and   the second   was an M-class flare peaking at about MJD 60048.6806. We removed them manually from the GTIs (the C-class flare from MJD 60048.53535 to MJD 60048.60033 and the M-class flare from  MJD 60048.67506 to MJD 60048.73744).

We found some relevant features in the energy spectrum below 4\,keV, most likely due to spectral features unaccounted for in the NICER ARF.
Because of the source's high count rate, these features become apparent in the spectral modeling. We therefore accounted for calibration artifacts owing to imperfections in modeling the dead layer of the silicon detector at the Si-K edge, and of the concentrator mirror surface roughness at the Au-M edges, which affects the $\approx$2.2–3.5 keV range.\footnote{\url{https://heasarc.gsfc.nasa.gov/docs/nicer/analysis_threads/arf-rmf/}} We froze the energy of the edges at their best-fit value, as reported in Table~\ref{tab:spectrum}.

\subsection{\INTEGRAL\ \IBIS}
\label{sec:integral}

\INTEGRAL\ \IBIS \citep{Winkler2003} observed the region of \gx from March 21 to 22 and from April 13 to 15, 2023, responding to a request by the \IXPE team. The data for this source are publicly available and were reduced for the imager IBIS \citep{Ubertini2003} and the ISGRI detector \citep{Lebrun2003}.
We used the MMODA\footnote{\url{https://www.astro.unige.ch/mmoda/}} platform that empowers the Off-Line Science Analysis (OSA) version 11.2 distributed by the ISDC \citep{Courvoisier2003} with the most recent calibration files that are continuously pushed in the instrument characteristic repository.
We first built a mosaicked image of all individual pointings that constitute the standard dithering strategy of observation for IBIS/ISGRI in the 28--40 keV energy range.
These images were used to make the catalog of detected sources with a signal-to-noise ratio higher than 7.
Using these catalogs, we extracted light curves with 1000\,s time bins and spectra in
256 standard channels for IBIS/ISGRI, these were grouped in ten equally spaced logarithmic  channels between 28 and 150\,keV.
We also accounted for a systematic uncertainty of the spectra at the 1.5\% level. 
The equivalent on-axis exposures of the IBIS/ISGRI spectra
are 40 and 85~ks, respectively, after correction for dead time and vignetting.
The products are available at the \INTEGRAL Product gallery.\footnote{\url{https://www.astro.unige.ch/mmoda/gallery/astrophysical-entity/gx-5-1}}

\subsection{\ATCA}
\label{sec:atca}

The Australia Telescope Compact Array (ATCA) observed \gx\ on March 24 and April 14, 2023. On March 24, the telescope observed 
with the array in its 750C configuration.\footnote{\url{https://www.narrabri.atnf.csiro.au/operations/array_configurations/configurations.html}} 
Observations taken on April 14 were carried out 
with the array in a relatively compact H214 configuration, 
in combination with an isolated antenna
located 6\,km from the array core, which was also included in our analysis. 
For both observations the data were recorded simultaneously at central frequencies of 5.5\,GHz and 9.0\,GHz, with 2\,GHz of bandwidth at each frequency.

We used PKS 1934$-$638 for bandpass and flux density calibration. PKS 1934$-$638 was also used to solve for the antenna leakages (D-terms) for the polarization calibration. The nearby source B1817$-$254 was used for gain calibration and to calibrate the PA using the  Common Astronomy Software Applications for radio astronomy (\textsc{casa}, version 5.1.2; \citealt{2022PASP..134k4501C}) \texttt{atcapolhelpers.py} task \texttt{qufromgain}.\footnote{ {\url{https://github.com/radio-astro}}} Calibration and imaging followed standard procedures within  \textsc{casa}. When imaging, we used a Briggs robust parameter of 0 to balance sensitivity and resolution \citep{1995PhDT.......238B}, and to  suppress the effects from some bright diffuse emission within the field.

For our March 24 observations, fitting for a point source in the image plane, we detected \gx\ at a flux density of $960 \pm 19$\,$\mu$Jy at 5.5\,GHz and $810 \pm 11\,\mu$Jy at 9\,GHz coincident with the previously reported radio position \citep[e.g.,][]{Berendsen2000,2007A&A...469..807L}. These detections correspond to a radio energy spectral index of $-0.37 \pm 0.09$. 
The Stokes $Q$ and $U$ values were measured at the position of the peak source flux density (Stokes $I$). 
No significant linearly polarized (LP) emission was detected at either frequency. Measuring the root mean square of the image noise in a $50\arcsec \times 50\arcsec$ region over the source position (taken as 1$\sigma)$, provides 3$\sigma$ upper limits on the polarized intensity $\sqrt{Q^2 + U^2}$ of $58\,\mu$Jy\,beam$^{-1}$ at 5.5\,GHz and $48\,\mu$Jy\,beam$^{-1}$ at 9\,GHz. 
These correspond to a 3$\sigma$ upper limit on the PD of 6.1\% at 5.5\,GHz and 5.9\% at 9\,GHz. 
Stacking the two frequencies to maximize the sensitivity also yields a nondetection of linearly polarized emission, with a 3$\sigma$ upper limit of 4.2\% (centered at 7.25\,GHz).

On April 14, following the same calibration and imaging procedure, we measured the flux density of \gx\ to be $750 \pm 50\,\mu$Jy and $620 \pm 40\,\mu$Jy at 5.5 and 9\,GHz, respectively. These detections correspond to a radio spectral index of $-0.4 \pm 0.1$. We note that due to the more compact array configuration for this epoch, and the presence of diffuse emission in the field, we imaged with a strictly uniform weighting scheme (setting the Briggs robust parameter to $-2$), reducing the impact of diffuse emission in the field on our resultant images. The compact configuration coupled with a shorter exposure time  resulted in a higher noise level in the images. At 5.5\,GHz, we do not detect any linearly polarized emission, with a 3$\sigma$ upper limit  on the PD of 12.5\%. Similarly, at 9\,GHz we measure a 3$\sigma$ upper limit on the PD of 20\%. Stacking the two frequencies places a  3$\sigma$ upper limit on the PD of 8\% at 7.25\,GHz. 
We   note that during the final $\sim$30\,min of the observation, some linear polarization was detected   close to the source position, but only at 9\,GHz. 
Due to the nondetection at 5.5\,GHz and the short and sudden nature of this emission, we attribute it to radio frequency interference and not an astrophysical event.

\subsection{VLT VISIR}\label{sec:visir}
Mid-IR observations of the field of \gx were made with the European Southern Observatory's Very Large Telescope (VLT) on March 28 and 31, 2023, under   program 110.2448 (PI: D. Russell). 
The VLT Imager and Spectrometer for the mid-Infrared \citep[VISIR;][]{Lagage2004} instrument on the VLT was used in small-field imaging mode. Four filters ($M$-band, $J8.9$, $B10.7$, and $B11.7$) were used, with central wavelengths of 4.67, 8.70, 10.64, and 11.51 $\mu$m, respectively. For each observation, the integration time on source was composed of a number of nodding cycles, chopping and nodding between source and sky. The total observing time was usually almost twice the integration time.

Observations of standard stars were made on the same nights as the target, in the same filters (photometric standards HD137744, HD169916, HD130157, and HD145897 were observed, all at airmass 1.0--1.1). Conditions were clear on both nights. All data (target and standard stars) were reduced using the VISIR pipeline in the \textsc{gasgano} environment.\footnote{\url{https://www.eso.org/sci/software/gasgano.html}} Raw images from the chop--nod cycle were recombined. Photometry was performed on the combined images using \texttt{PHOT} in \textsc{IRAF}.\footnote{IRAF is distributed by the National Optical Astronomy Observatory, which is operated by the Association of Universities for Research in Astronomy, Inc., under cooperative agreement with the National Science Foundation.} For the B10.7 filter, two standard stars were observed on each night, just before and after each observation of \gx, to check for stability. We found that the counts-to-flux ratio values from these standards agree to a level of 6.4\%, which we adopt as the systematic error of all flux measurements. The estimated counts-to-flux ratio in each filter was used to convert count rates (or upper limits) of \gx to flux densities. 

On March 28, 2023 (MJD 60031.37), the B10.7 filter was used only, and we derive a 3$\sigma$ flux density upper limit of 3.27\,mJy at 10.64 $\mu$m (the airmass was 1.05--1.10). On March 31, 2023  (MJD 60034.34), \gx was detected in $M$-band and $J8.9$, with flux densities of $4.2 \pm 1.4$\,mJy at 4.67 $\mu$m and $10.0 \pm 2.3$ mJy at 8.70 $\mu$m, respectively (at airmass 1.07--1.17). The significance of the detection was 5.9$\sigma$ in both filters. The errors on the fluxes incorporate the statistical error on each detection, and the systematic error from the standard stars, in quadrature. The source was not detected in the $B10.7$ and $B11.7$ filters on March 31, 2023, with flux upper limits that were less constraining than on March 28,  2023.

\gx lies in a crowded region of the Galactic plane, with several stars detected within 5\arcsec\ of the source. The near-IR counterpart was confirmed through photometry and spectroscopy \citep[named star 513;][]{Jonker2000,Bandyopadhyay2003}, and its coordinates agree with the radio and X-ray position. We are confident that the source we detect with VISIR is indeed \gx because star 503 from \citet{Jonker2000} and \citet{Bandyopadhyay2003} is also detected (at a low significance of 3.1--3.3$\sigma$) at the correct coordinates. This star is estimated from Fig. 1 of \citet{Jonker2000} to lie 4\farcs3 to the southeast of \gx. In the $J8.9$ VISIR image the detected source is measured to be 4\farcs28 to the southeast of the position of the detected \gx, as expected. The flux density of star 503 is $1.4 \pm 0.5$ mJy in $M$-band and $4.5 \pm 2.2$ mJy in $J8.9$.

\begin{figure}
\centering
\includegraphics[width=0.95\columnwidth]{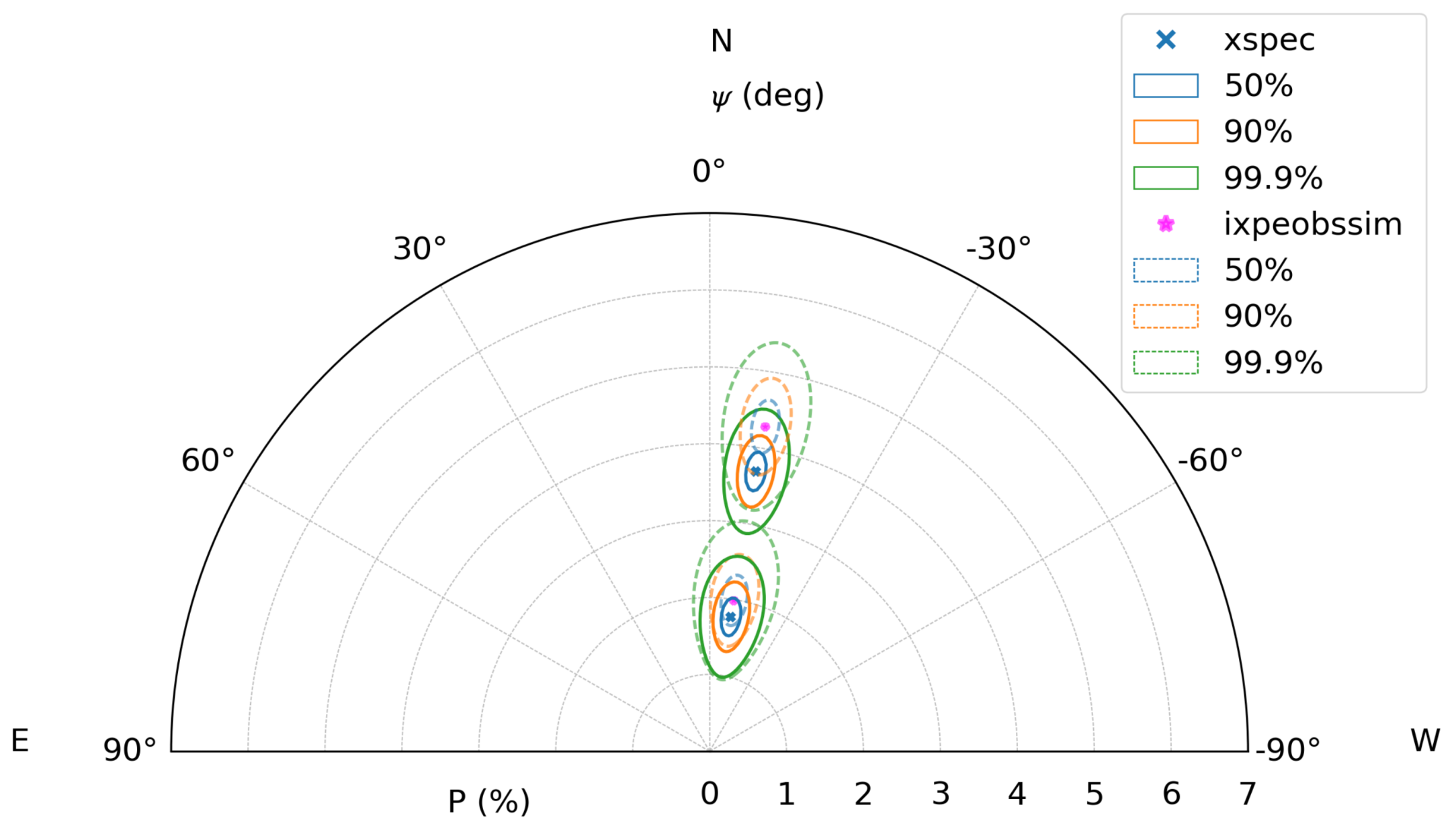}
\caption{Polarization of \gx measured by \IXPE in the 2--8~keV energy range during Obs.~1 and Obs.~2 obtained with \textsc{xspec} and \textsc{ixpeobssim}. 
The contours are computed for the two parameters of interest at 50\%, 90\%, and 99.9\% confidence levels.
}
\label{fig:totPDPA}
\end{figure}

\subsection{REM and LCO}\label{sec:rem} 

Optical (SDSS $griz$ filters) and near-infrared (NIR; 2MASS $H$-band) observations of \gx were acquired with the robotic 60\,cm Rapid Eye Mount (REM; \citealt{Zerbi2001}; \citealt{Covino2004}) telescope on March 22 and on April 14, 2023 (see Table~\ref{tab:obslog}).
Strictly simultaneous observations were obtained in all bands; on March 22, a series of 150 exposures of  30 s  each were acquired in H-band (dithering was applied), and 20 exposures of 300 s  in each of the optical bands; on April 14, a series of 225 exposures of 30 s  were acquired in H-band (dithering was applied), and 30 exposures of 300 s  in each of the optical bands.

Optical observations were also acquired with the 1\,m and 2\,m telescopes of the Las Cumbres Observatory (LCO) network, using the $i'$ and $Y$ filters. Observations with the 2\,m telescope (Siding Spring - Faulkes Telescope South) were performed on 2023-03-22T16:33:07; 2023-03-29T15:50:11; 2023-03-30T15:46:39; and 2023-03-31T15:43:07 (300s integration in all epochs and bands), and with the 1\,m telescopes (at the locations of Cerro Tololo, Siding Spring, and Sutherland) on 2023-03-21T07:09:39; 2023-03-22T01:01:36; 2023-03-22T08:11:35, and 2023-04-14T05:35:19, T08:11:37, T15:11:30, and T23:23:55 (300s integration in all epochs and bands).

The optical images were bias- and flat-field corrected using standard procedures; the contribution of the sky in the NIR images was evaluated by performing a median of the dithered images five-by-five, and was then subtracted from each image.
In all bands and epochs, all the reduced images were then aligned and averaged in order to increase the signal-to-noise ratio. 
Aperture photometry was performed using \texttt{PHOT} in \textsc{IRAF}.
Flux-calibration was performed against a group of six stars with magnitudes tabulated in the 2MASS catalog\footnote{\url{https://irsa.ipac.caltech.edu/Missions/2mass.html}}  and six stars from the PanSTARRS catalog.\footnote{\url{https://catalogs.mast.stsci.edu/panstarrs/}}

Due to the very high extinction of the source ($N_{\rm H}=4.93\times 10^{22}\, \rm cm^{-2}$ and $4.88\times 10^{22}\, \rm cm^{-2}$, which translates\footnote{See Sect. \ref{sec:distance} and \citet{Foight2016}.} into $A_V=17.2$ mag and 17~mag in Obs.~1 and Obs.~2, respectively) and to the combination of the low spatial resolution of our images and the crowded field of the source, \gx is not detected in any of the optical and NIR images acquired. A blend of our target with at least one of the nearby stars can be detected at very low significance in the averaged $H$-band image, at a position consistent with the proposed optical and NIR counterpart of the source (\citealt{Jonker2000}, star 513). However, this detection is not significant enough to extract a flux for the blend.

\begin{figure*}
\centering
\includegraphics[width=0.47\textwidth]{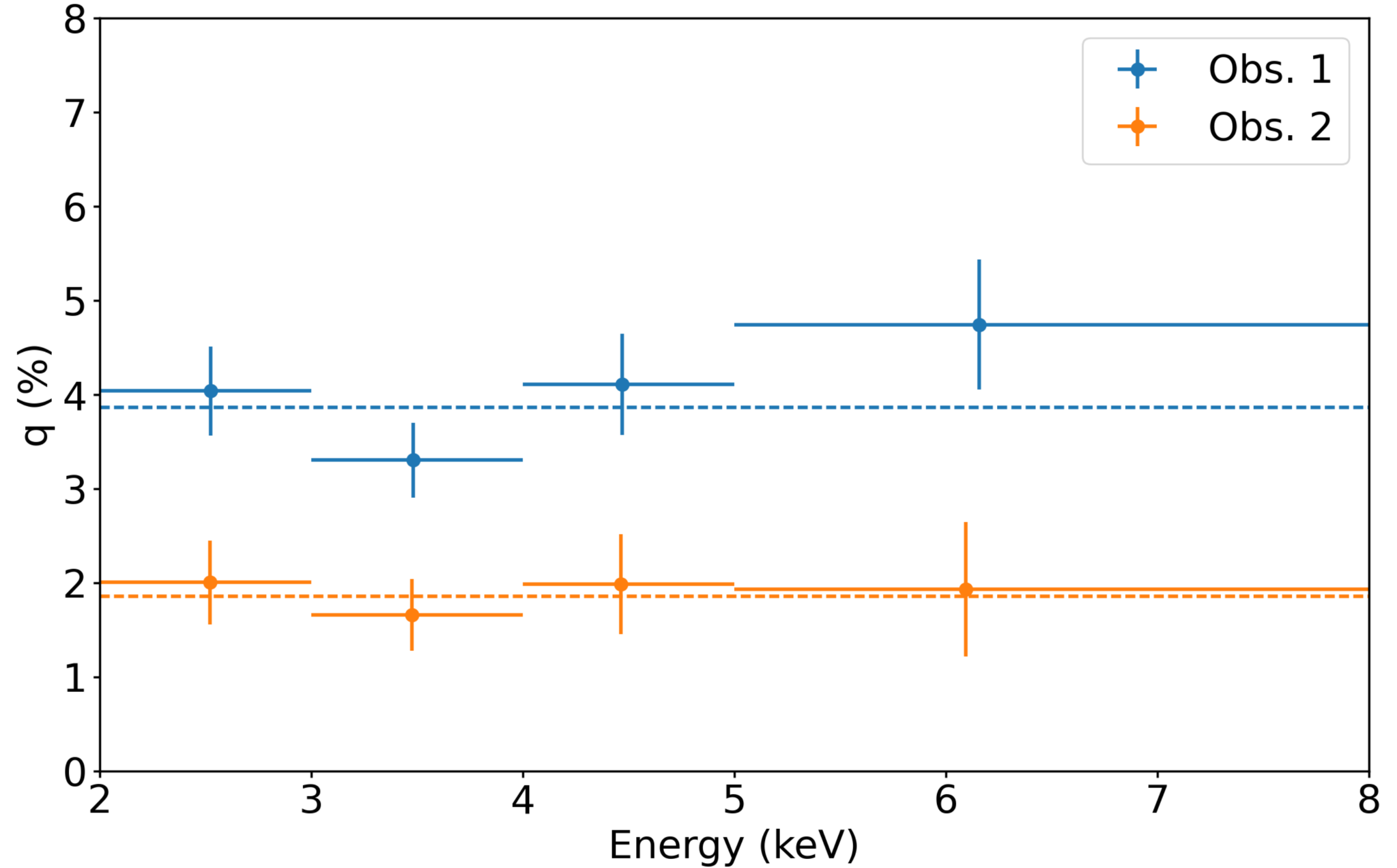}
\includegraphics[width=0.48\textwidth]{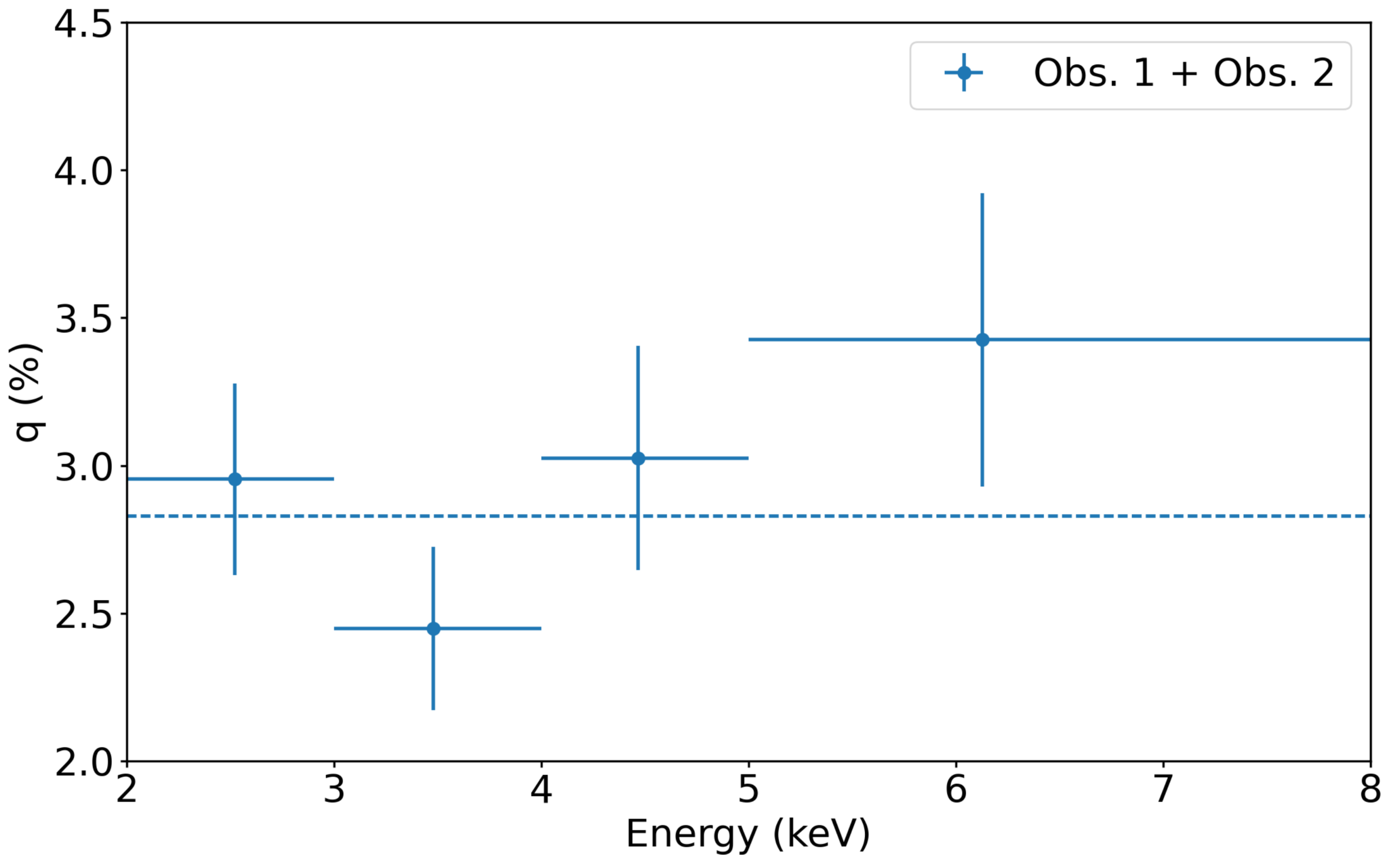}

\includegraphics[width=0.47\textwidth]{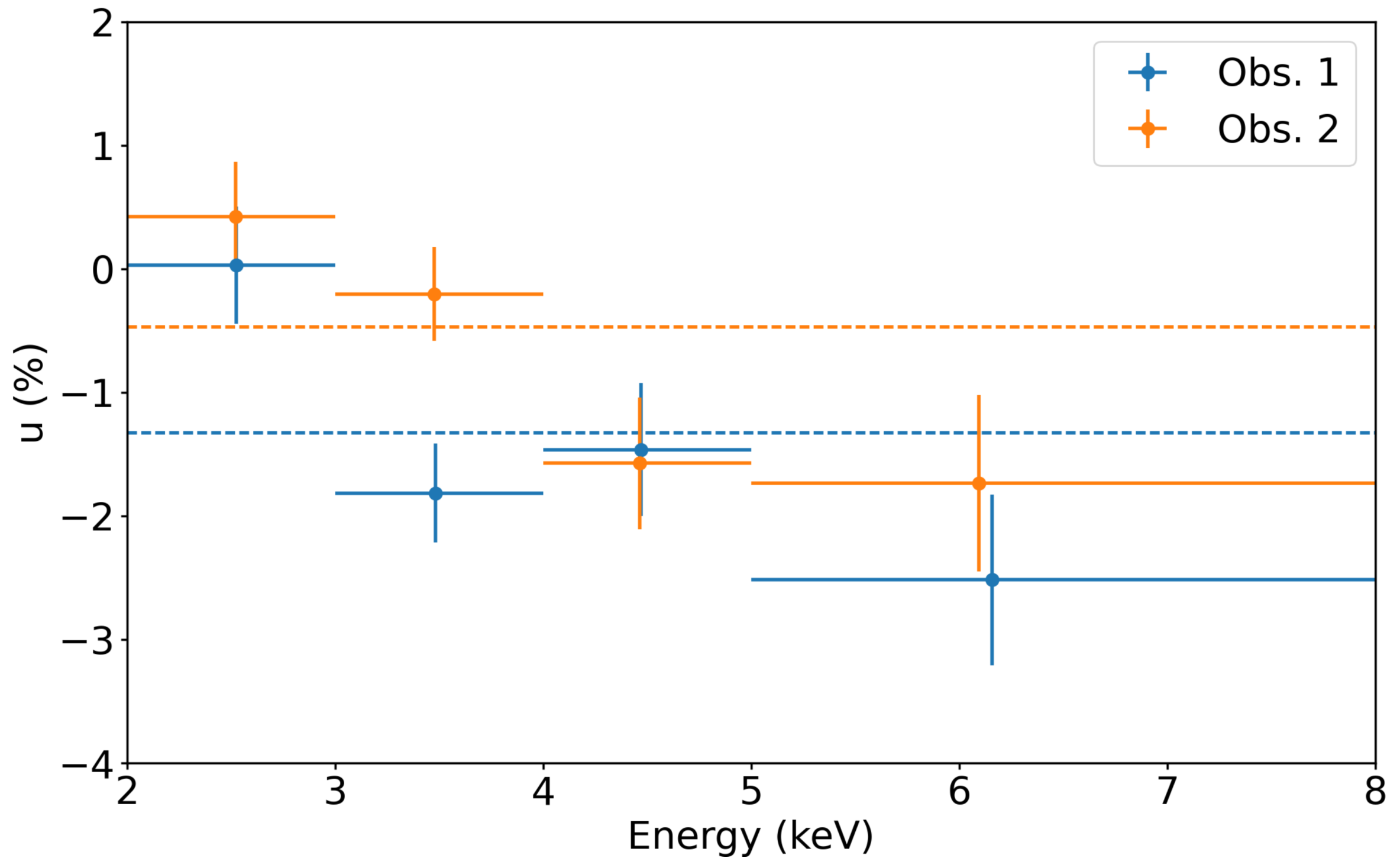}
\includegraphics[width=0.48\textwidth]{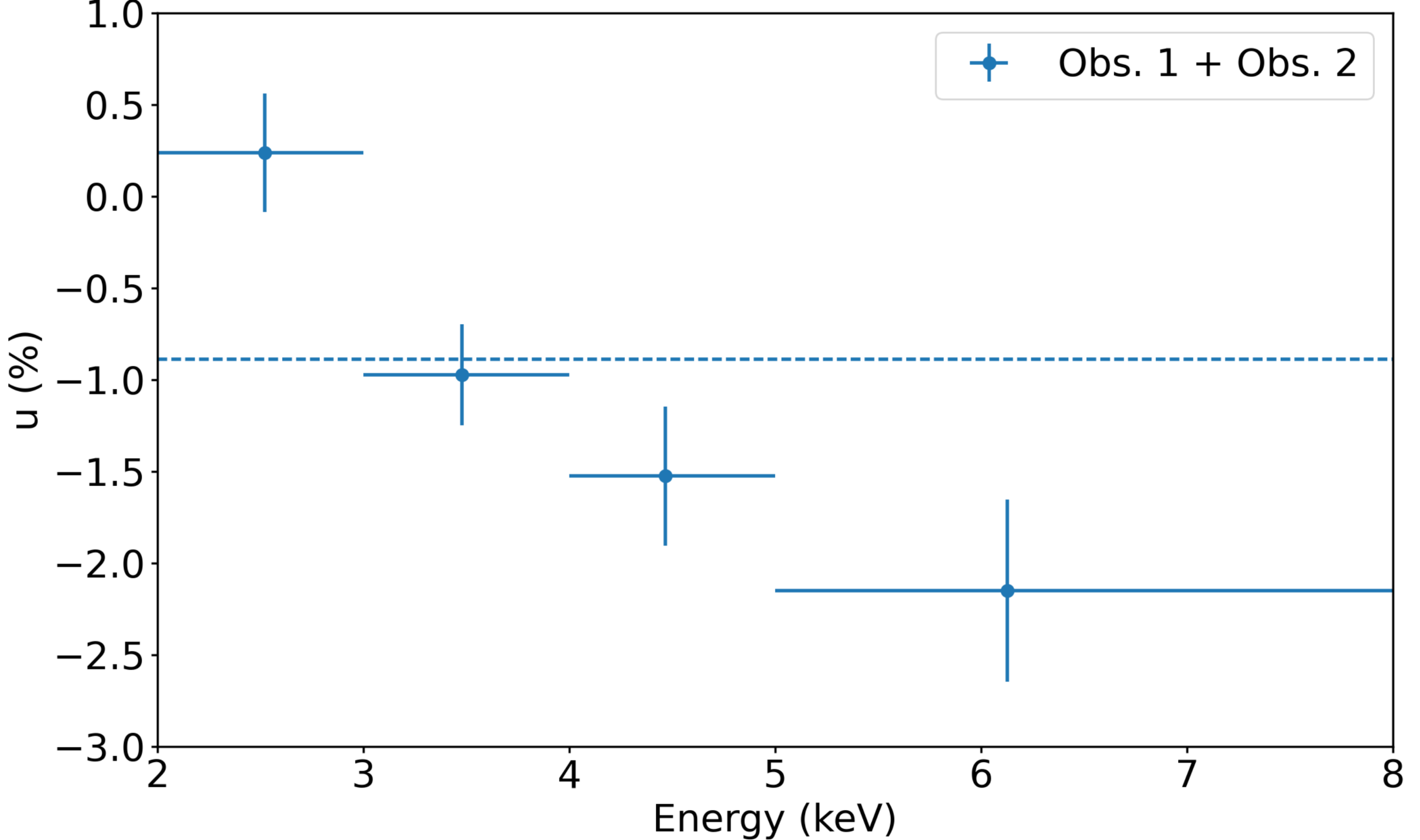}
\caption{Stokes parameters as a function of energy for \IXPE Obs.~1 and Obs.~2 (left panels) and for the combined data set (right panels)  obtained with \textsc{ixpeobssim}. The normalized Stokes $q$ parameter is compatible with a constant value in each   of the \IXPE observations, whereas the Stokes $u$ parameter is not (see Table~\ref{tab:stokesconst}). 
This behavior of the Stokes parameters is consistent with a variation in  PA (see Fig.~\ref{fig:contouranglerot}).} 
\label{fig:stokes1and2andtot}
\end{figure*}

\begin{table} 
\centering
\caption{Polarization in the 2--8~keV band (see~Fig. \ref{fig:totPDPA}) and in four energy bins estimated with \textsc{ixpeobssim} for Obs.~1 and Obs.~2.}
\begin{tabular}{c l c c  }
\hline
\hline
Energy (keV) &Parameters & Obs. 1 & Obs. 2   \\
\hline
\multirow{2}*{2--8} &PD ($\%$)&
$4.3\pm0.3$  &
$2.0\pm 0.3$  \\
&PA (deg)&
$-9.7\pm2.0$  &
$-9.2\pm 4.0$  \\
\hline
\multirow{2}*{2--3} &PD ($\%$)&
$4.0\pm0.5$  &
$2.0\pm 0.4$  \\
&PA (deg)&
$0.2\pm3.4$  &
$6.0\pm 6.2$  \\
\hline
\multirow{2}*{3--4} &PD ($\%$)&
$3.8\pm0.4$  &
$1.7\pm 0.3$  \\
&PA (deg)&
$-14.4\pm3.8$  &
$-3.5\pm 6.5$  \\
\hline
\multirow{2}*{4--5} &PD ($\%$)&
$4.4\pm0.5$  &
$2.5\pm 0.5$  \\
&PA (deg)&
$-9.8\pm3.5$  &
$-19.2\pm 6.0$  \\
\hline
\multirow{2}*{5--8} &PD ($\%$)&
$5.4\pm0.7$  &
$2.6\pm 0.7$  \\
&PA (deg)&
$-14.0\pm3.7$  &
$-21.0\pm 7.9$  \\
\hline
\end{tabular}
\tablefoot{The errors are at 68\% confidence level.}\label{tab:papdixpeobssim}
\end{table}

We estimate the following $3\sigma$ upper limits (only the most constraining ones per epoch are quoted): $H=13.84$, $Y=19.08$, $z=18.28$, $i'=21.79$ (LCO), $r'=20.22$, $g'=20.45$ for Obs 1; $H=14.00$, $Y=18.86$, $z=18.35$, $i'=21.22$, $r=20.46$, $g=20.50$ for Obs 2.
These upper limits are consistent with the $H$-band magnitude of the proposed NIR counterpart reported by \citet{Jonker2000} (Star 513; $H=14.1\pm 0.2$).

\section{Results}
\label{sec:results}

\subsection{Polarimetric model-independent analysis}\label{sec:modelindep}

The model-independent analysis of the X-ray polarization with {\tt ixpeobssim} (see Table~\ref{tab:papdixpeobssim}) in the 2--8~keV energy band gives the PD of  $4.3\%\pm0.3\%$ and $2.0\%\pm 0.3\%$ in Obs.~1 and Obs.~2, respectively (with 1$\sigma$ errors). The corresponding PAs are $-9\fdg7\pm2\fdg0$ and $-9\fdg2\pm 4\fdg0$.
We see a significantly larger  PD in Obs. 1 compared to Obs. 2. However, the PA is compatible with being constant during the two observations. The contour plots of the \textsc{ixpeobssim} analysis in the 2--8~keV energy range are reported in Fig.~\ref{fig:totPDPA} (dashed lines).

\begin{table} 
\centering
\caption{Results of the Stokes parameters fit with a constant value for Obs.~1, Obs.~2, and the combined data set.}
\scriptsize
\begin{tabular}{ccccc}
\hline
\hline
Stokes  & Best-fit  & Obs. 1 & Obs. 2 &  Obs. 1 + Obs. 2   \\ 
parameter & parameters & & & \\
\hline
& constant (\%)& $3.86\pm 0.25$ & $1.86\pm0.24$&  $2.83\pm 0.17$  \\
$q$ & $\chi^2_{\nu}$ & 1.31 & 0.15 &   1.26  \\
& $\alpha$ (\%) & 26.8 & 93.0 &   28.7  \\
\hline
& constant (\%)& $-1.32\pm0.25$ & $-0.74\pm0.24$ & $-0.89\pm0.17$    \\
$u$ & $\chi^2_{\nu}$& 4.26 & 4.00 &  7.16   \\
& $\alpha$ (\%) & 0.52 & 0.74 & 0.01    \\
\hline
\end{tabular}
\tablefoot{The reduced $\chi^2$ of the fit is $\chi^2_{\nu}$. The  significance level of the $\chi^2$ value is $\alpha$. The number of degrees of freedom is 3.}  
\label{tab:stokesconst}
\end{table}

The behavior of polarization as a function of energy is reported in Table \ref{tab:papdixpeobssim}. The energy dependence of the PD is essentially the same in the two observation, whereas the PA varies from positive to negative values. A proper assessment of this behavior requires the use of Stokes parameters. The left panels of Fig. \ref{fig:stokes1and2andtot} show the Stokes parameters as a function of energy of the two  observations separately. 
For each observation, the Stokes $q$ parameters are consistent with being constant in energy, albeit at different values between the first and second observations.
On the other hand, the Stokes $u$ parameters are not compatible with a constant (see Obs.~1 and Obs.~2 columns of Table~\ref{tab:stokesconst}). This requires the variation in  PA.

\begin{figure}
\centering
\includegraphics[width=0.49\textwidth]{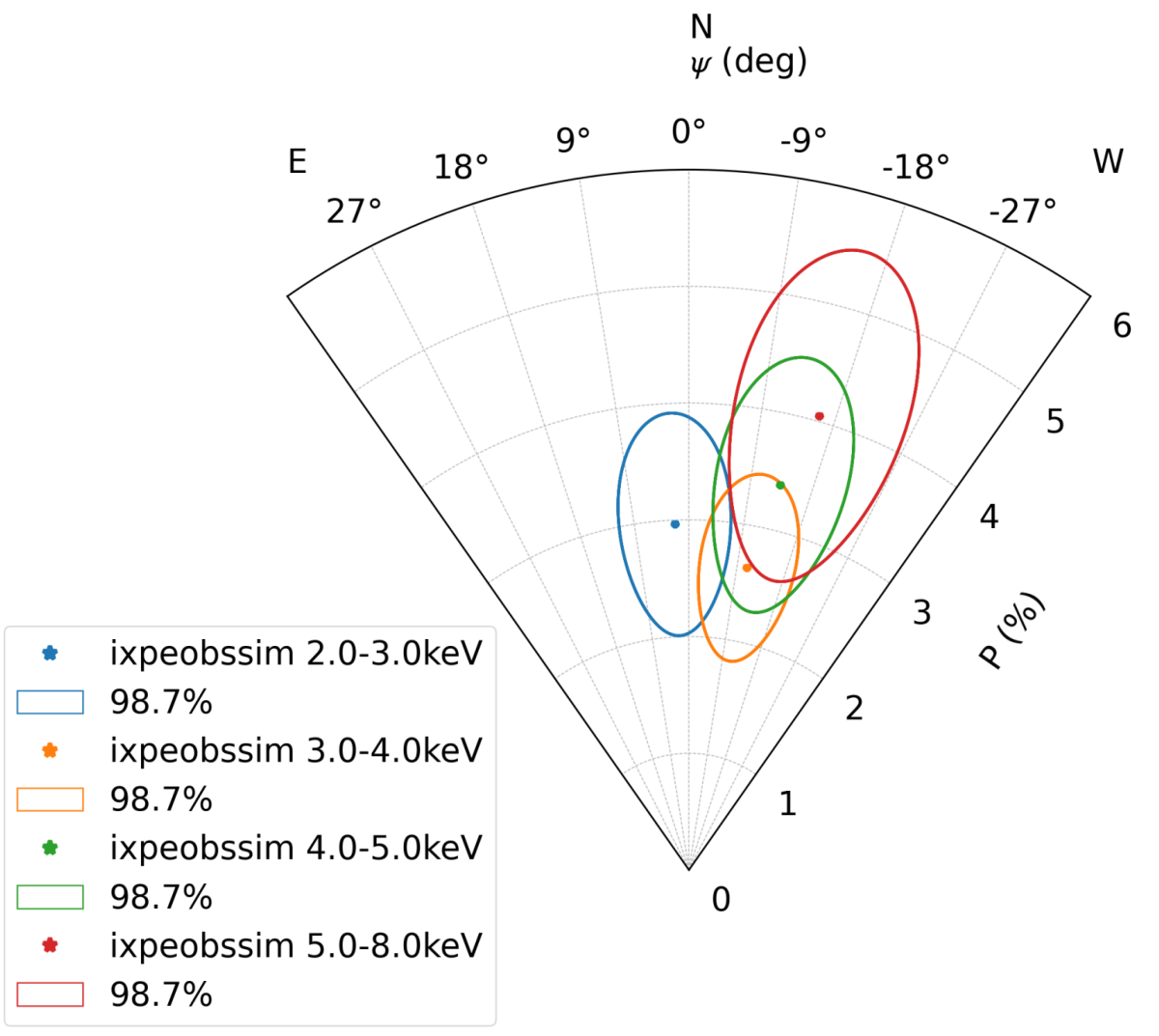}
 
\vspace*{0.5cm}
\includegraphics[width=0.48\textwidth]{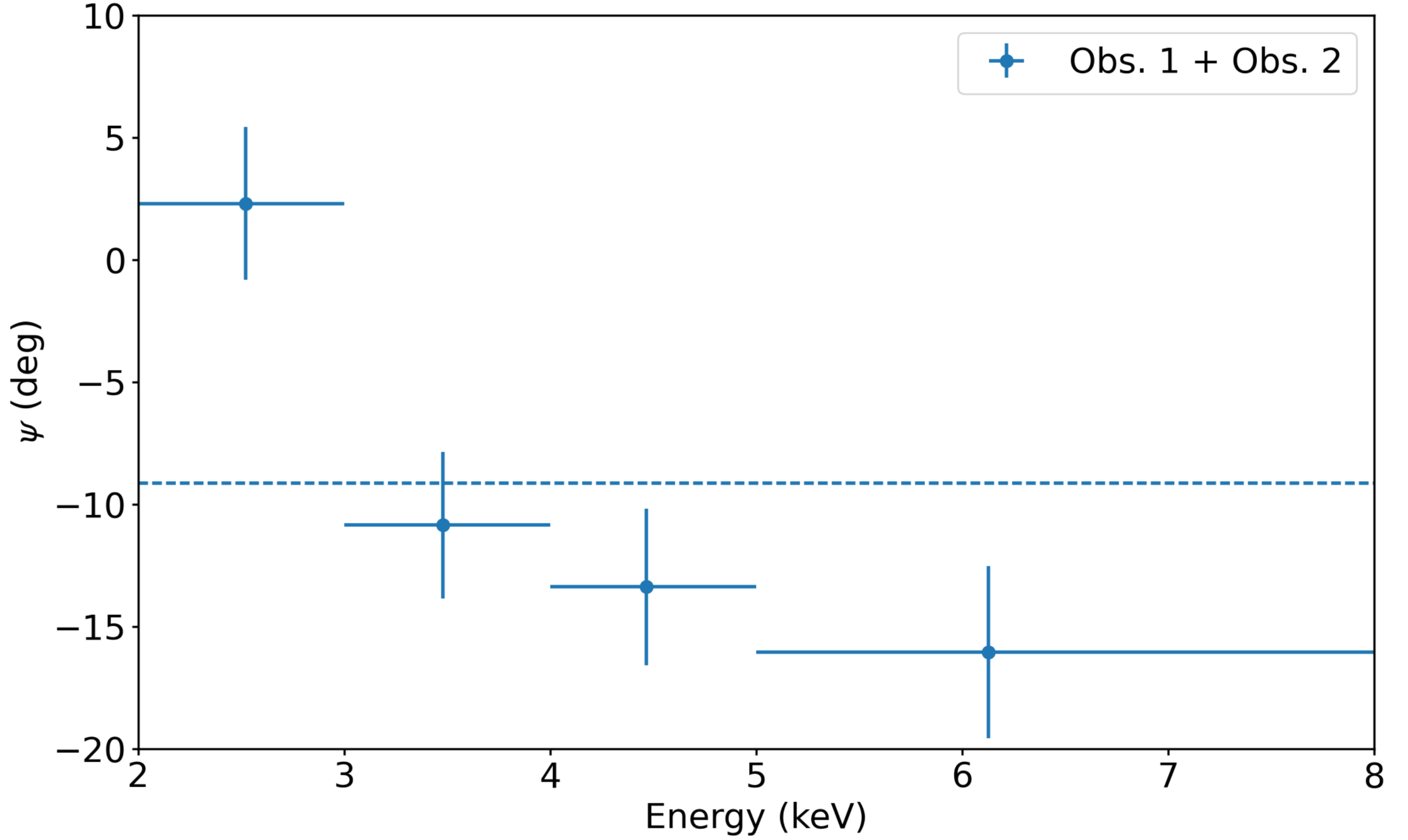}
\caption{Polarization contour plot (top panel) and PA (bottom panel) as a function of energy of the combined \IXPE data set (Obs.~1 and Obs.~2). }
\label{fig:contouranglerot}
\end{figure}

On the basis of the assumption that the geometry and the physical process producing polarization are the same in the two observations, we also calculated the Stokes parameters of the \IXPE observations combined (Obs. 1 and Obs. 2) to improve the statistics. 
While the resulting normalized Stokes parameter $q$   remains  compatible with a constant value, the Stokes $u$ is even   farther from being a constant with the reduced $\chi^2$ values for $\nu=3$ degrees of freedom being $\chi^2_{\nu}=1.26$ and $\chi^2_{\nu}=7.16$ for $q$ and $u$, respectively. 
As anticipated by the separate analysis of the two observations, this behavior of $q$ and $u$ implies a variation in  PA.
Such a variation of about 20\degr\  is highlighted in Fig.~ \ref{fig:contouranglerot}. In the top panel the contour plots on the PD--PA plane for the \IXPE whole observation are shown. Polarization in the 2--3~keV energy bin is not compatible with that in the 5--8~keV bin with a probability of $\sim 98.7\%$. 
In the bottom panel the  variation in PA with energy is shown together with the fit by a constant giving a value of $-9\fdg1\pm1\fdg6$ with   $\chi^2_{\nu}=6.44$ for $\nu=3$ corresponding to a significance level $\alpha=0.02\%$.

\begin{figure*}
\centering
\includegraphics[width=0.9\textwidth]{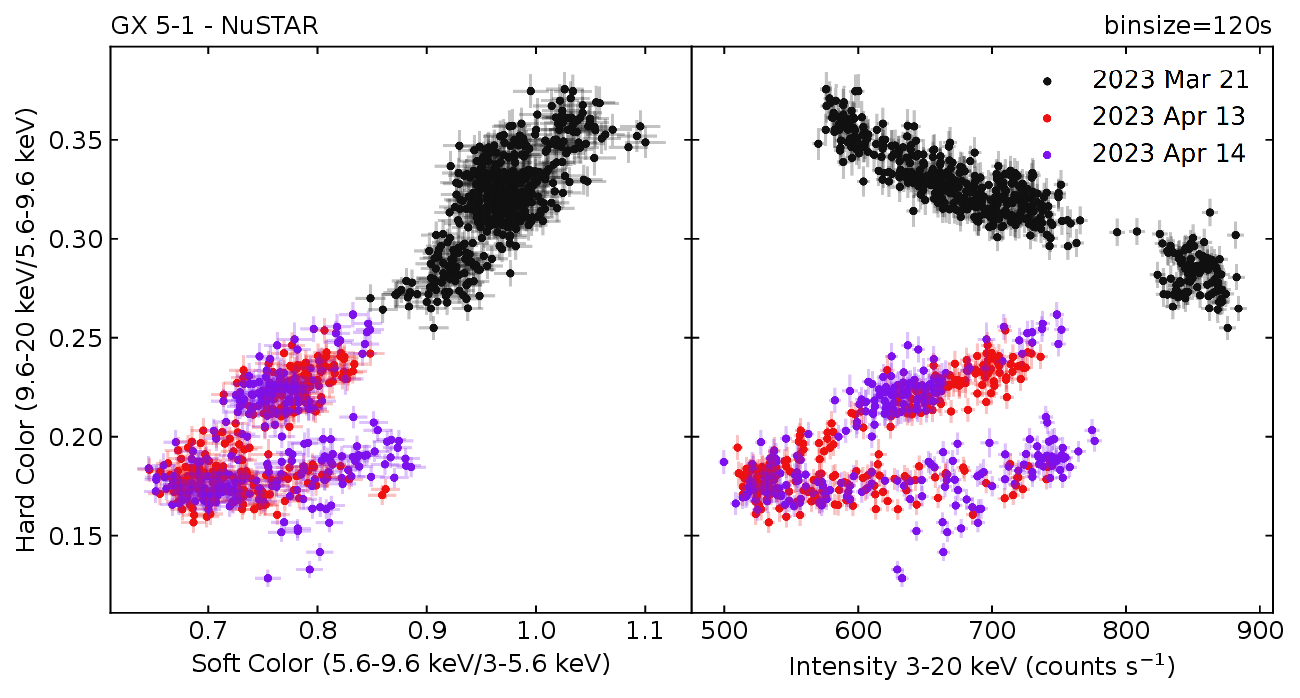}
\caption{\Nustar CCD and HID of \gx contemporary to \IXPE observation. During \IXPE Obs.~1 the source was in the HB, while during Obs.~2 it was in the NB-FB. The observing days  March 21, April 13,  and April 14, 2023, are shown in black, red, and purple,   respectively.}
\label{fig:ccdnustar}
\end{figure*}

\subsection{X-ray spectral analysis}\label{sec:spectralanalysis}

The \NICER, \Nustar, and \IBIS light curves contemporary to \IXPE observations are shown in Fig.~\ref{fig:lc}. 
The time-resolved CCD and hardness-intensity diagram (HID) for all the observations with \IXPE, \NICER, and \Nustar show \gx moving along the complete Z-track from March 21 to April 14. \Nustar CCD and HID, obtained from the GTIs contemporary to the \IXPE observations (see Fig.~\ref{fig:ccdnustar}), highlight clearly the Z shape, disentangling also the NB with respect to the FB, due to the wide energy band from 3 to 20~keV used to construct the CCD colors.
During \IXPE Obs.~1 the source was in the HB, whereas it was in the NB-FB during Obs.~2, which we  checked  by constructing the  CCD and HID from the \IXPE data of the two observations.
From March 24 to 25 (about 3 days after the end of Obs.~1), when ATCA was observing, \NICER detected \gx moving from the HB-NB corner toward the NB.

\begin{table} 
\centering
\caption{Best-fit parameters of \gx spectral model from \NICER, \Nustar, and \IBIS data simultaneous to \IXPE observation.}
\scriptsize
\begin{tabular}{llcc}
\hline \hline
Components & Parameters &
        Obs. 1 & Obs. 2   \\
\hline
    {\tt edge} & $E$ (keV) &
          \multicolumn{2}{c}{1.82148 (frozen)}\\
            & $\tau$ &
         $0.172^{+0.028}_{-0.016}$ &
          $0.152\pm 0.017$ \\
    {\tt edge} & $E$ (keV) &
         \multicolumn{2}{c}{1.95197 (frozen)}\\
            & $\tau$ &
         $0.051\pm0.016$ &
  $0.484\pm0.016$\\
    {\tt edge} & $E$ (keV) &
        \multicolumn{2}{c}{2.28003 (frozen)}\\
            & $\tau$ &
         $0.037\pm0.014$ &
          $0.025\pm0.014$\\
    {\tt edge} & $E$ (keV) &
         \multicolumn{2}{c}{2.44444   (frozen)}\\
            & $\tau$ &
         $0.050\pm0.013$ &
          $0.038\pm 0.013$ \\
    {\tt edge} & $E$ (keV) &
         \multicolumn{2}{c}{3.16139   (frozen)}\\
            & $\tau$ &
         $0.020^{+0.007}_{-0.008}$ &
          $0.012\pm 0.007$ \\ 
          \hline
   \multicolumn{4}{c}{Continuum paramters}\\
    {\tt tbabs}  & $N_{\rm H}$ ($10^{22}$  cm$^{-2}$)&
         $4.93^{+0.12}_{-0.06}  $ &
         $4.88\pm0.04$   \\
{\tt diskbb} &  $kT$  (keV) &
        $0.95\pm0.07$ &
        $1.20\pm0.03$ \\
        & $R_{\rm in}\sqrt{\cos{\theta}}$  (km)\tablefootmark{a}&
        $25^{+2}_{-3}$  & $19.6^{+0.9}_{-0.8}$  \\
    {\tt thcomp} & $\Gamma$  &
        $2.35^{+0.13}_{-0.49}$ &
        $<2.1$  \\
          & $kT_{\rm e}$ (keV)  &
          $2.99^{+0.02}_{-0.10} $ &
           $3.08^{+0.37}_{-0.11}$\\
          &  $f$   &
          $0.99^{+0.01}_{-0.36}$&
          $0.032^{+0.08}_{-0.005}$ \\
{\tt bbodyrad}   &  $kT_{\rm bb}$  (keV)&
 $1.27^{+0.26}_{-0.11}$ & $1.68^{+0.03}_{-0.04}$   \\
             & $R_{\rm bb}$  (km)\tablefootmark{a}&
             $19^{+1}_{-6}$  
             & $9.3^{+0.6}_{-0.5}$\\
{\tt expabs}   & $E_{\rm cut}$ (keV) &
[= $kT_{\rm bb}$] & --\\
{\tt powerlaw}   & $\Gamma $ &       $2.62^{+0.49}_{-0.92}$  &
                      --  \\
& norm &
$0.45^{+2.3}_{-0.43}$  &
                      --  \\
\hline 
\multicolumn{4}{c}{Cross-calibration constants}\\
{\tt const}   &  $C_{\rm NICER}$   &
$1.107\pm0.003$ &
$1.034\pm 0.004$   \\
&   $C_{NuSTAR,\rm  FPMA}$  &
\multicolumn{2}{c}{1 (frozen)}\\
&  $C_{NuSTAR,\rm  FPMB}$   &
$1.0136\pm0.0013$ &
$0.9908\pm 0.0014$ \\
& $C_{\rm IBIS/ISGRI}$ &
$0.66\pm0.08$ & $1.29\pm0.03$\\
&  $\chi^2$/d.o.f.
&  421/451
&  338/365   \\
\hline
& $\tau_{\tt thcomp}$\tablefootmark{b} & $8.7^{+2.5}_{-0.7}$ & $>10.0$\\
& $f_{2-8\,\rm keV}$\tablefootmark{c}   & 2.3 & 2.4\\
& $f_{2-5\,\rm keV}$\tablefootmark{c}  & 1.5 & 1.7\\
& $f_{5-8\,\rm keV}$\tablefootmark{c}   & 0.8 & 0.7\\
& $f_{\rm ({\tt thcomp*bbodyrad})}$ / $f_{2-8\,\rm keV}$ & 61$\%$ & 33$\%$\\
\hline
\end{tabular}
\tablefoot{The errors are at 90\% confidence level. The edges reported in the top section of the table   refer  only to the \NICER energy spectrum.
\tablefoottext{a}{Radii are estimated assuming the distance to the source of 7.6~kpc (see Sect.~\ref{sec:distance}).}  
\tablefoottext{b}{The optical depth $\tau_{\tt thcomp}$ comes from Eq.~(14) in \citet{Zdziarski2020}.}
\tablefoottext{c}{The unabsorbed flux is measured in units of 10$^{-8}$~erg~cm$^{-2}$~s$^{-1}$.}}
\label{tab:spectrum} \end{table}

\begin{figure*}
\centering
\includegraphics[width=0.495\textwidth]{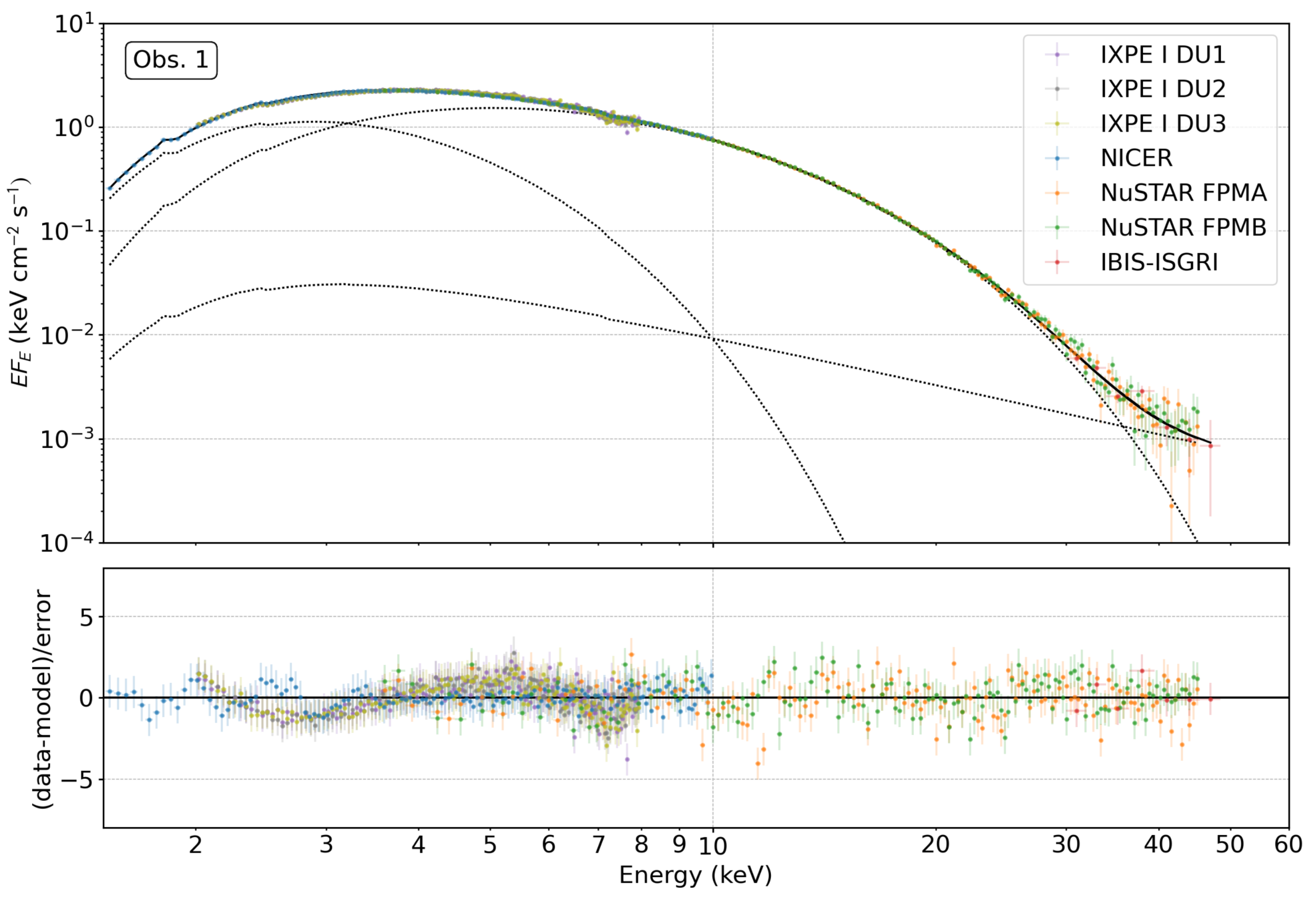}
\includegraphics[width=0.495\textwidth]{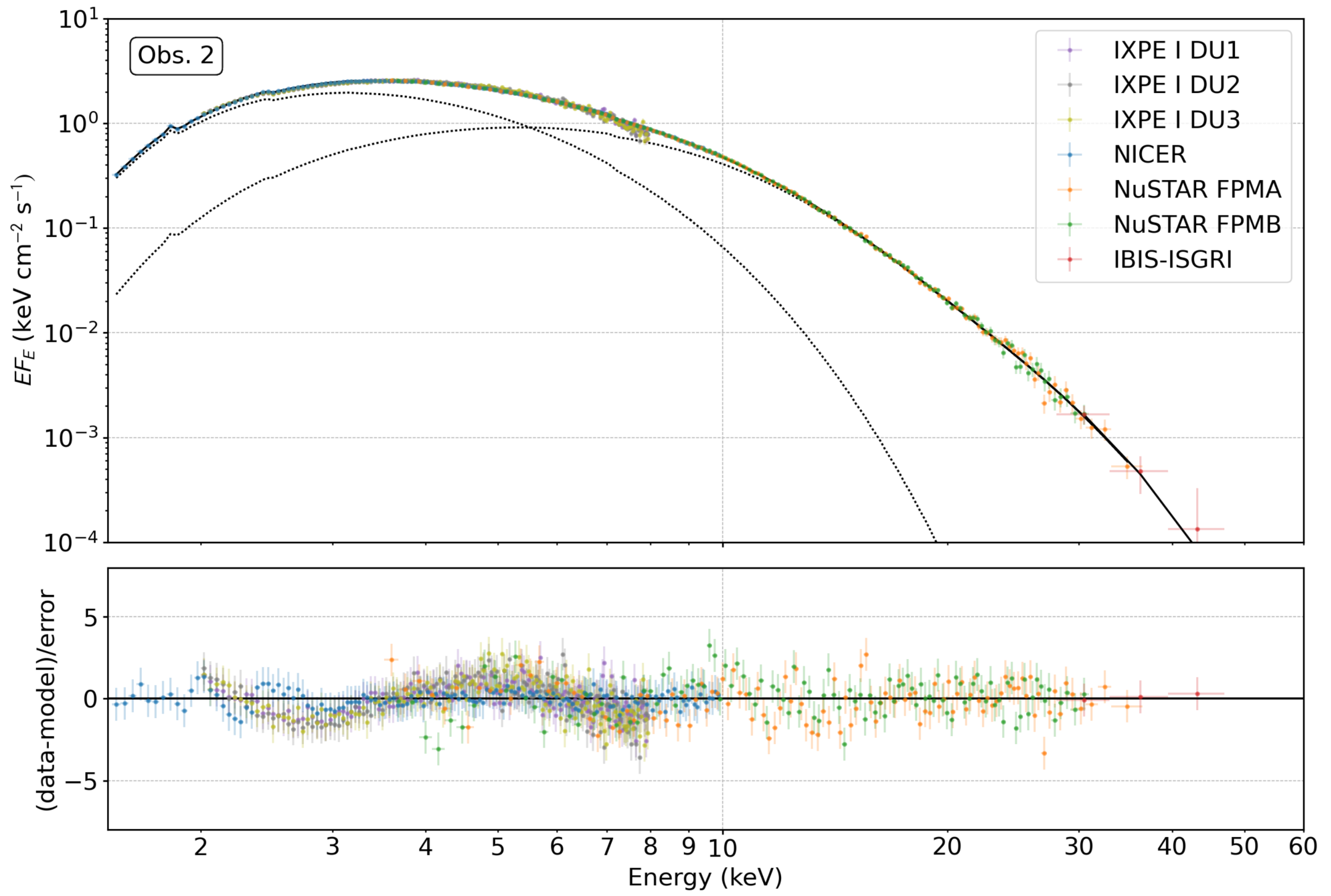}
\caption{Deconvolved energy spectrum of \gx. The top panels show the deconvolved \NICER (1.5--10~keV), \Nustar (3.5--45~keV for Obs.~1 and 3.5--35~keV for Obs.~2), and \IBIS (28--50~keV) spectra simultaneous with both \IXPE observations, obtained from the best-fit model  in Table \ref{tab:spectrum}. The \IXPE (2--8~keV) energy spectrum is overplotted on the frozen model. The bottom panels show the residuals between data and model in units of~$\sigma$.}
\label{fig:spectralfit}
\end{figure*}

We fit the \NICER, \Nustar, and \IBIS data  of Obs.~1 and
Obs.~2 and present the results in Fig.~\ref{fig:spectralfit} (where the \IXPE energy spectrum is overplotted to the frozen model) and Table~\ref{tab:spectrum}.
The better fits we obtained are based on the following spectral models for Obs.~1 and Obs.~2: 
\begin{description}
\item {\tt tbabs*(diskbb+expabs*powerlaw+thcomp*bbodyrad)}, 
\item  {\tt tbabs*(diskbb+thcomp*bbodyrad)},
\end{description}
respectively. 

We included the normalizing cross-calibration multiplicative factors for the \NICER, \IBIS, and \Nustar FPMA (frozen at unity) and FPMB telescopes.
A \texttt{tbabs} \citep{Wilms2000} multiplicative model component was used to take into account the low-energy absorption due to the interstellar medium. We used the abundances and cross-section tables according to \citet{Wilms2000} and \citet{Verner1996} (\texttt{wilm, vern} in \xspec).
We modeled the \gx energy spectra of both the observations with a multicolor disk and a harder boundary layer (BL) or spreading layer (SL) emission \citep[e.g.,][]{Popham01, Revnivtsev13}. In both observations, the convolution model component {\tt thcomp} \citep{Zdziarski2020} is used to represent Comptonized emission from the BL--SL  \citep[e.g.,][]{disalvo02,Farinelli09} modeled with the {\tt bbodyrad} component.
The $f$ parameter of {\tt thcomp} represents the fraction of Comptonized seed photons. In Obs. 1 this parameter is $>0.63$ with the best-fit value equal to 0.99. Effectively, all photons from the BL--SL {\tt bbodyrad} are Comptonized. In Obs. 2 only a fraction between 2.7\% and 11.2\% (best-fit 3.2\%) of seed photons are Comptonized. It is worth noting that a higher Comptonization fraction corresponds to a higher PD  observed from the source.
During Obs.~1, when the source is on the HB, the energy spectrum is harder, and a high-energy excess was seen (but absent in Obs.~2). This excess is modeled with an additional hard tail represented by a power-law component with a low-energy exponential roll-off. The e-folding energy for the absorption of the exponential roll-off is set equal to the {\tt bbodyrad} $kT_{\rm bb}$ energy of the SL--BL. This parameter link comes from the assumption that the power-law emission originates from the seed distribution of the SL--BL blackbody. 
The BL and the inner disk are hotter in Obs.~2 than in Obs.~1, while the BL sphere-equivalent radius and inner disk radius are larger in Obs.~1.
In contrast to Obs.~1, the additional power-law component is not needed to model the spectrum of Obs.~2.

The softer disk component dominates up to $\sim 3.5$~keV in Obs.~1 and up to $\sim 5.5$~keV in Obs.~2 (see Fig.~\ref{fig:spectralfit}).
In the 2--8~keV band, the energy flux in the Comptonized component (including the contribution of the exponentially absorbed power-law component) accounts for $\sim 61\%$ of the total in Obs.~1. In Obs.~2 the energy flux of the Comptonized component drops to $\sim 33\%$ of the total in the same energy band.

\subsection{X-ray spectropolarimetric analysis}\label{sec:modeldependent}

The results of the spectropolarimetric analysis, including \IXPE, once the spectral model used to fit the data from the other observatories is frozen, is reported in Table~\ref{tab:gainfitixpe}.
The PD obtained with a {\tt polconst} multiplicative model component applied to the whole spectral model of \gx in Obs.~1 and Obs.~2 is $3.7\%\pm0.4\%$ and $1.8\%\pm 0.4\%$, respectively. The PA is around $-9\degr$ in both the observations.
The polarization contour plot in the 2--8~keV energy band obtained with \textsc{xspec} ({\tt polconst} model applied to the spectral model) is shown in Fig.~\ref{fig:totPDPA}. 
These contours are nearly identical to those obtained using \textsc{ixpeobssim}. 

\begin{table} 
\centering
\caption{Best-fit parameters of polarization analysis with \textsc{xspec}.}
\scriptsize
\begin{tabular}{ll c c  }
\hline
\hline
DU & Parameters &
Obs. 1 & Obs. 2   \\
& & &       \\
\hline
DU1 & $N_{\rm DU1}$ &
$0.796\pm0.006$ &
$0.798\pm0.004$ \\
     & gain slope &
$0.953^{+0.003}_{-0.03}$ &
$0.961\pm0.003$     \\
& gain offset (keV) &
$0.13\pm0.02$ &
$0.102\pm0.015$ \\
DU2 & $N_{\rm  DU2}$ &
$0.772\pm0.006$ &
$0.769\pm0.004$ \\
 & gain slope &
$0.952\pm0.003$ &
$0.960\pm0.003 $ \\
& gain offset (keV)&
$0.15\pm0.02$ &
$0.131\pm0.014$ \\
DU3 & $N_{\rm DU3}$ &
$0.734\pm0.005$ &
$0.736\pm 0.004$ \\
& gain slope &
$0.976\pm 0.004$ &
$0.966\pm 0.003$ \\
& gain offset (keV) &
$0.10\pm 0.02$ &
$0.114\pm 0.015$ \\
\hline
Components & Parameters &
        Obs. 1 & Obs. 2   \\
\hline
\multicolumn{4}{c}{{\tt polconst*(diskbb+[expabs*powerlaw]\tablefootmark{a}+thcomp*bbodyrad)}}\\
& PD ($\%$)&
$3.7\pm0.4$  &
$1.8\pm0.4$  \\
& PA (deg)&
$ -9\pm3$  &
$-9\pm6$  \\
&  $\chi^2$/d.o.f.
&  1806/1799
&  1733/1711  \\
\multicolumn{4}{c}{{\tt polconst$_1$*diskbb+polconst$_2$*([expabs*powerlaw]\tablefootmark{a}+thcomp*bbodyrad)}}\\
{\tt polconst$_1$}                    & PD ($\%$)&
$2.3\pm0.9$  &
$1.8\pm 0.9$  \\
& PA (deg)&
$21\pm11$  &
$14\pm15$  \\
{\tt polconst$_2$}
& PD ($\%$)&
$5.7\pm1.4$ & 
$4.3\pm 2.0$  \\
& PA (deg)&
$-16\pm20$  &
$-32\pm14$  \\
&  $\chi^2$/d.o.f.
&  1792/1806
&  1726/1718  \\
\hline
\end{tabular}
\tablefoot{The errors are at 90\% confidence level (see Table~\ref{tab:spectrum} for the corresponding spectral analysis). Polarization is computed in the 2--8~keV energy range. 
\tablefoottext{a}{\texttt{expabs*powerlaw} component was included only in Obs. 1.}}  
\label{tab:gainfitixpe}
\end{table}

\begin{figure}
\centering
\includegraphics[width=0.5\textwidth]{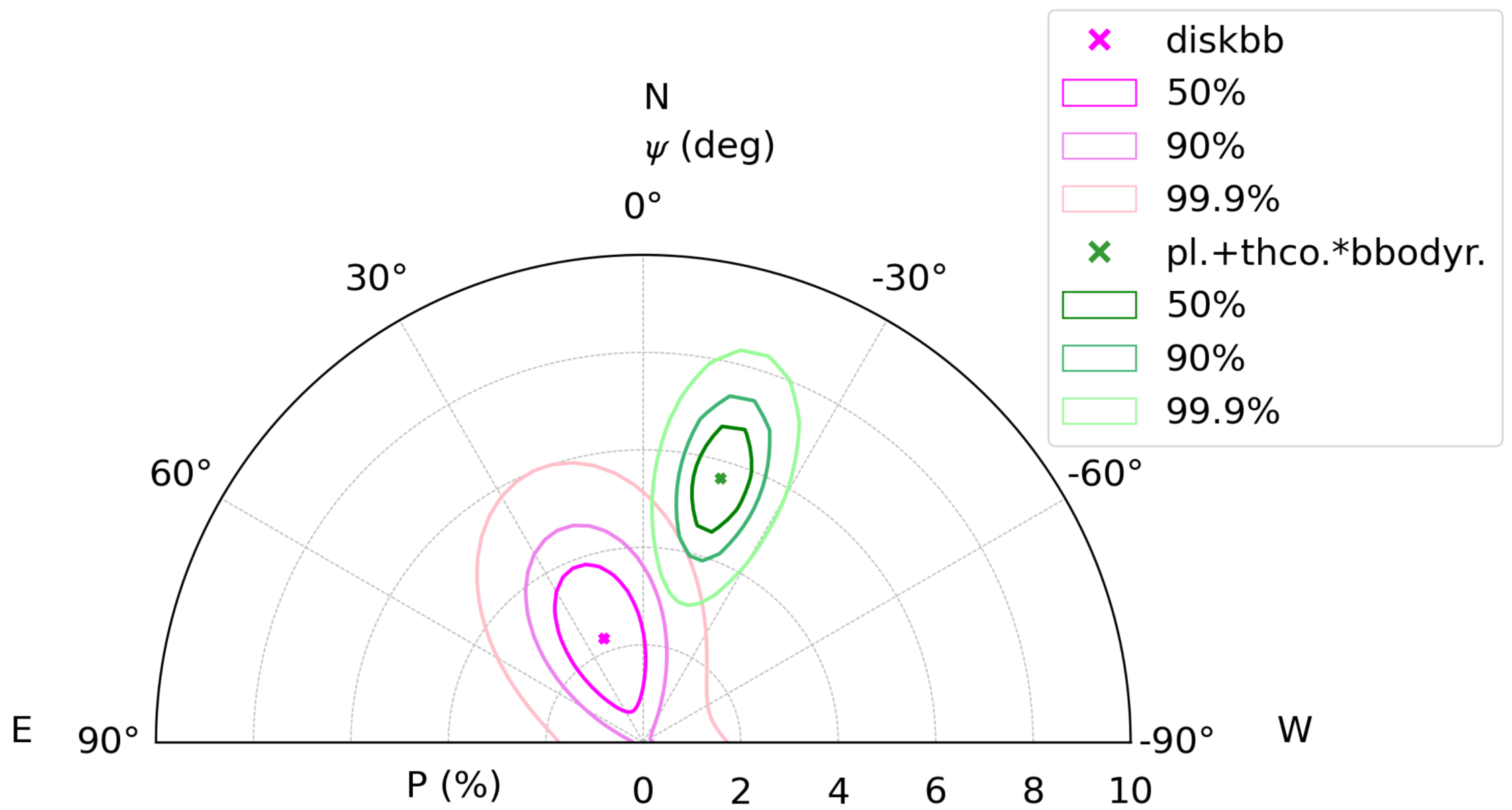}

\hspace*{-1cm}
\includegraphics[width=0.43\textwidth]{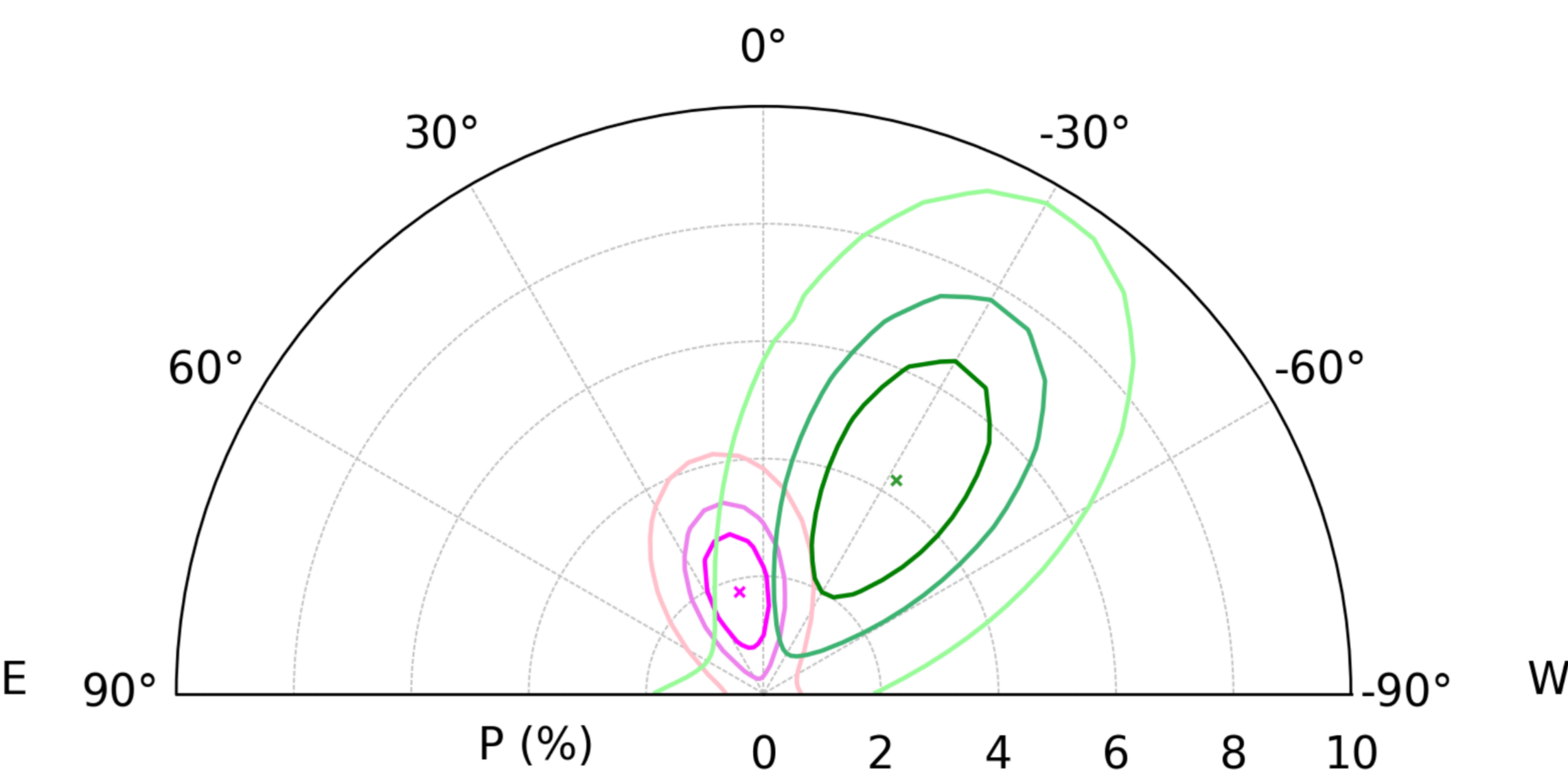}
\caption{Polarization contour plots from the spectropolarimetric analysis in the 2--8~keV energy range of Obs.~1 (top panel) and Obs.~2 (bottom panel). The {\tt polconst} polarimetric model is applied separately to the {\tt diskbb} component of both \IXPE observations and to the {\tt (expabs*powerlaw+thcomp*bbodyrad)} components as a whole. 
The {\tt expabs*powerlaw} component (labeled  {\tt pl.} in the  legend) is included only in Obs.~1. 
Contour plots are computed for the four parameters of interest, taking into account the fact that the polarization of the disk and Comptonization components are correlated in the simultaneous fit.}
\label{fig:modelspolconst}
\end{figure}

The spectropolarimetric analysis with \textsc{xspec} allows us to assign a polarization to the different spectral components. The result of this analysis is reported in Table~\ref{tab:gainfitixpe} and Fig.~\ref{fig:modelspolconst}. 
For Obs.~1 we assigned the same polarization for the {\tt expabs*powerlaw} and  {\tt thcomp*bbodyrad} components, assuming that the power law is just a continuation of the BL--SL component, with the power-law low-energy rollover $E_{\rm cut}$ being equal to the temperature of the BL--SL {\tt bbodyrad} $kT_{\rm bb}$.

\begin{figure*}
\centering
\includegraphics[width=0.48\textwidth]{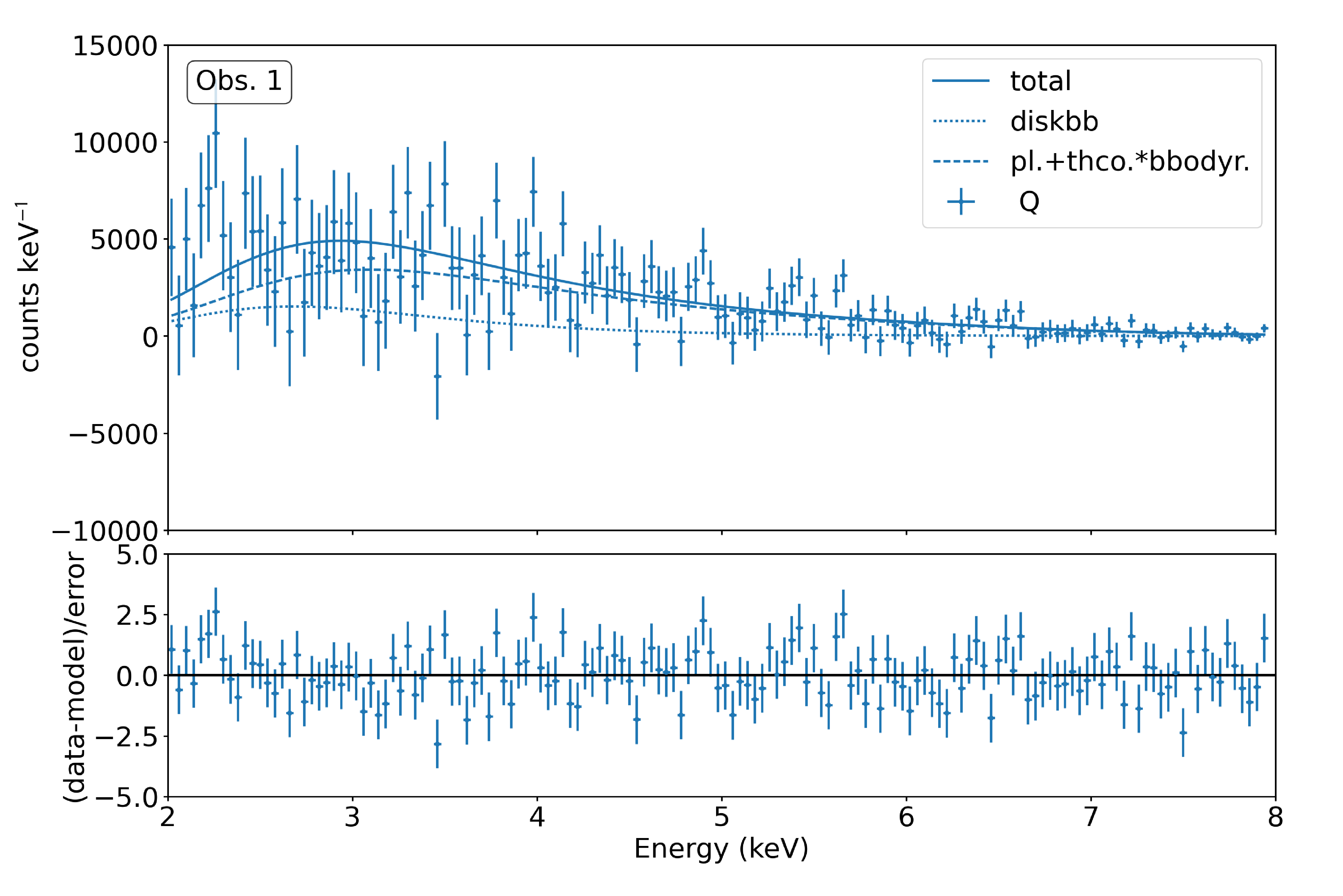}
\includegraphics[width=0.48\textwidth]{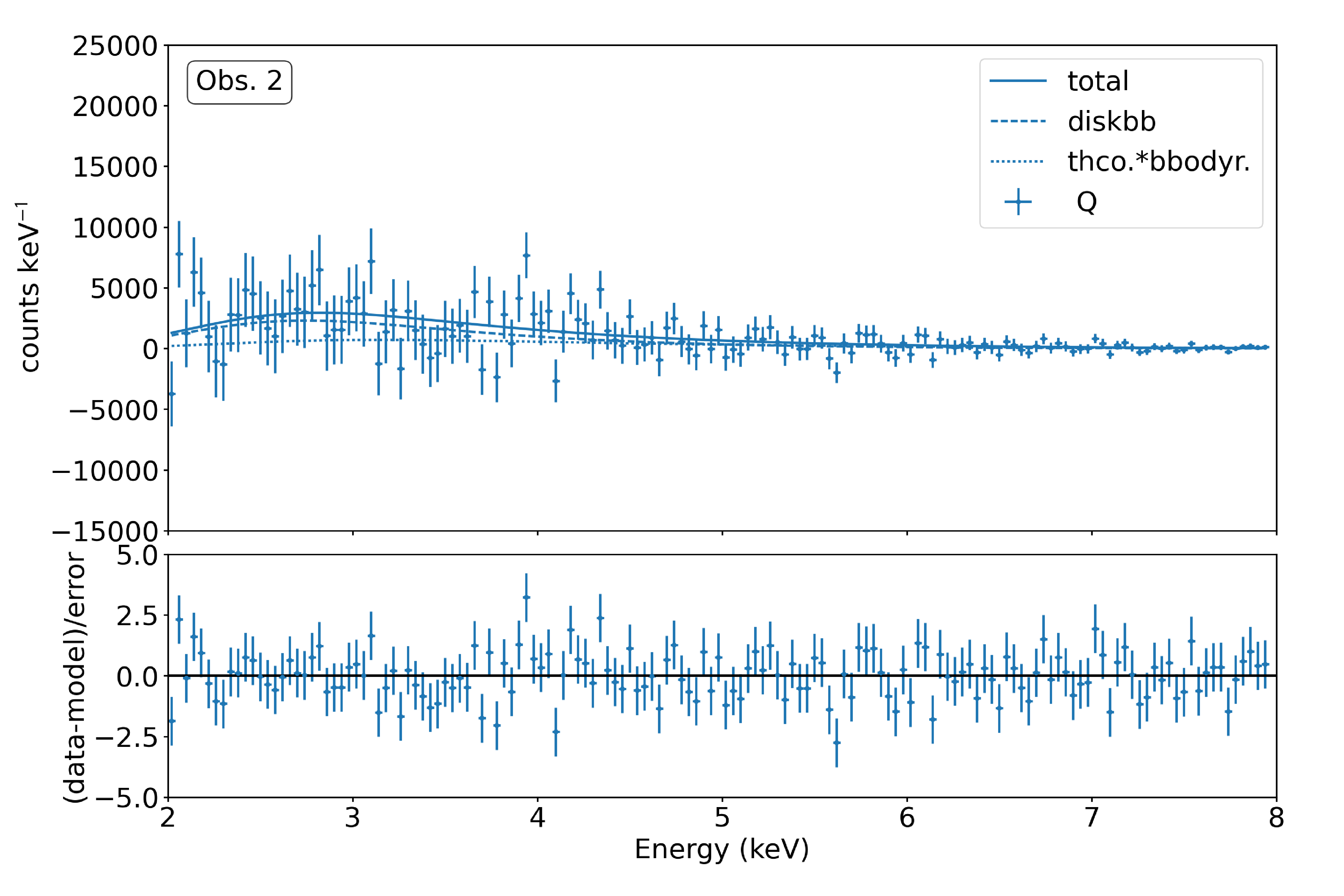}

\includegraphics[width=0.48\textwidth]{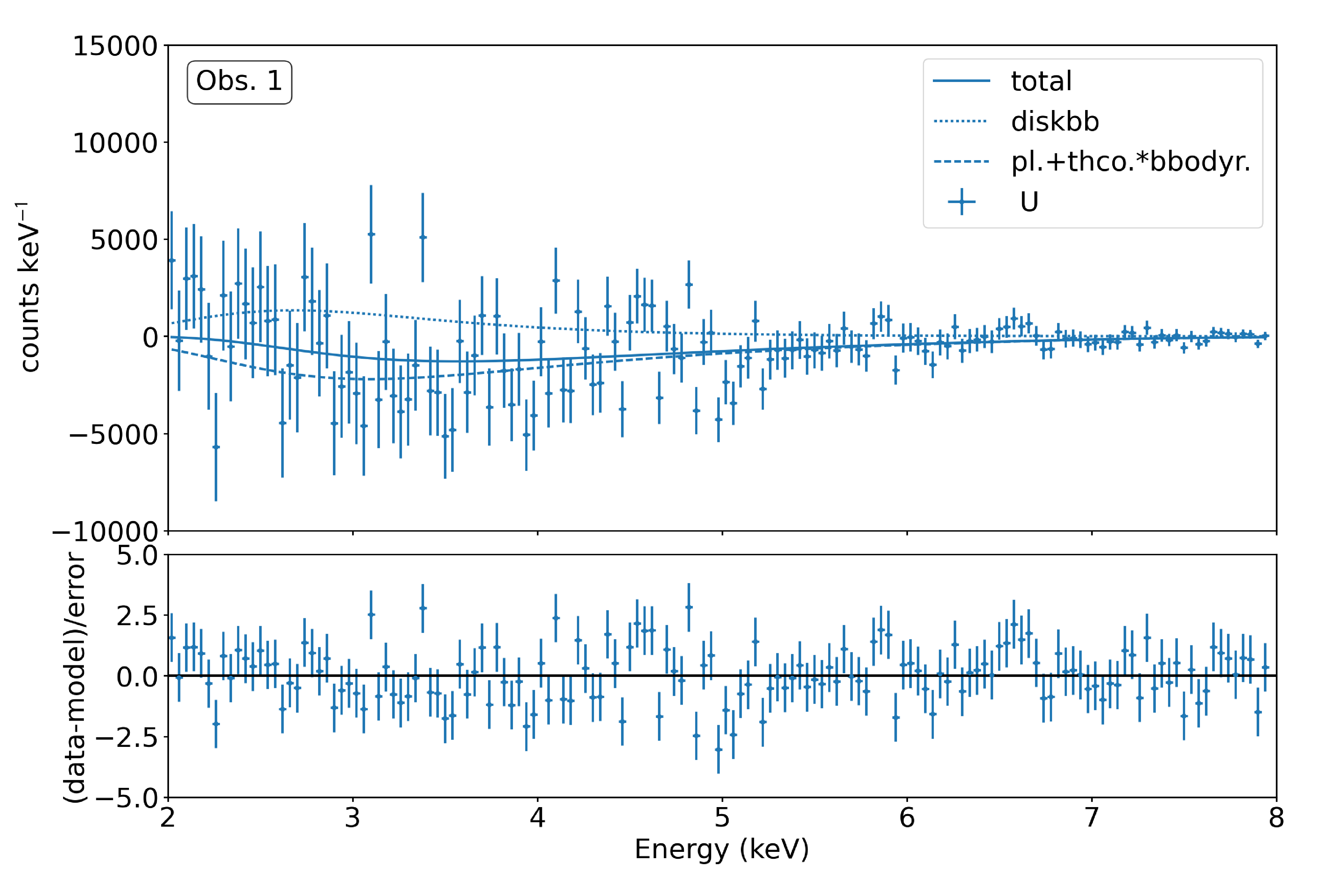}
\includegraphics[width=0.48\textwidth]{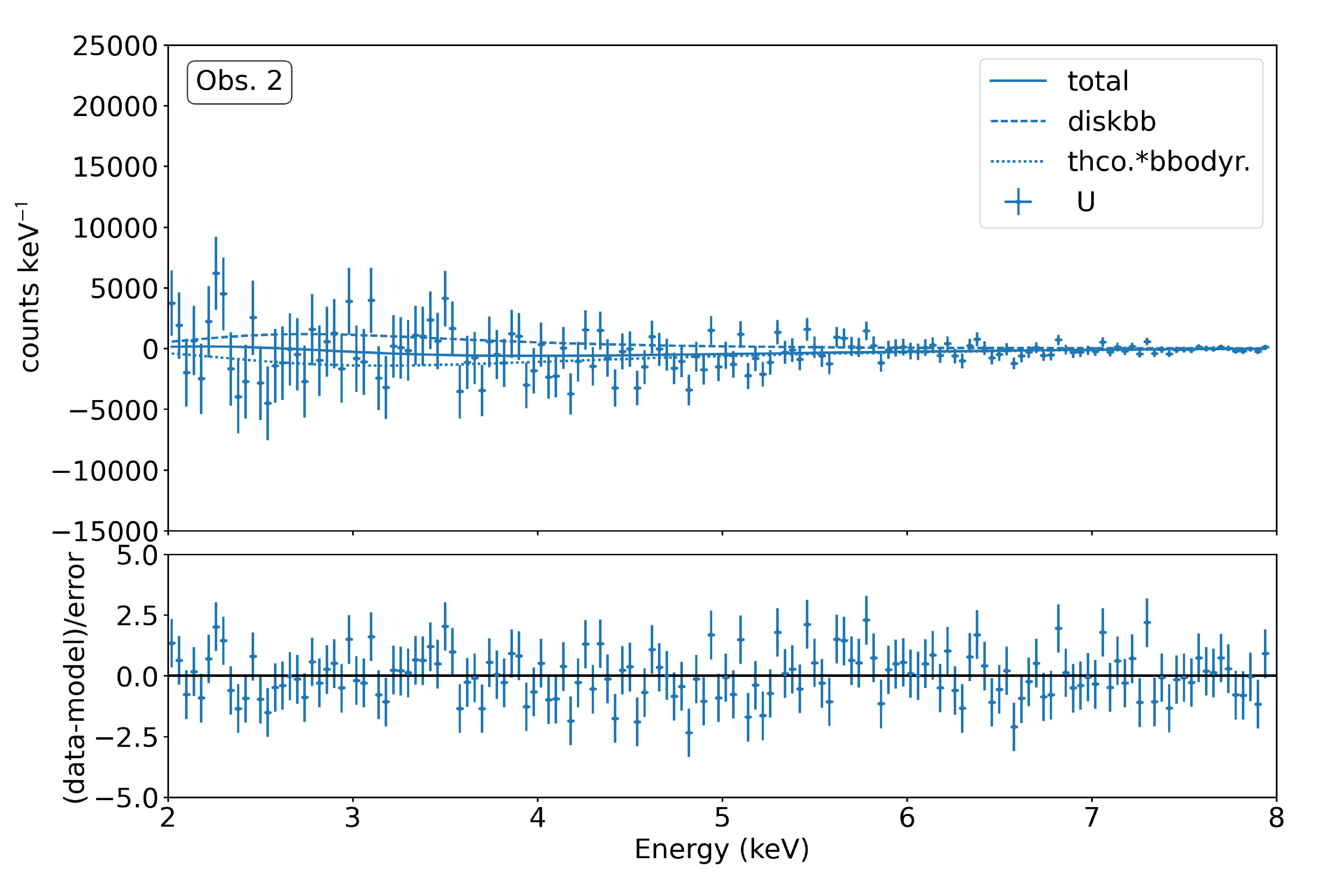}
\caption{Fit of the \IXPE Stokes parameters $Q$ and $U$  as a function of energy with the model comprising two {\tt polconst} components applied to the disk (dotted lines) and to the Comptonization (dashed lines) for Obs.~1 (left column) and Obs. 2 (right column).}
\label{fig:stokesfit}
\end{figure*}

The Comptonization component is well constrained at a 99.9\% confidence level only in Obs.~1, while the disk component is not constrained at 99\% in either observation (see Fig.~\ref{fig:modelspolconst}). Moreover,
in both the observations the polarization of each component has similar PAs, suggesting that the geometry responsible for the polarization is similar. Figure~\ref{fig:stokesfit} shows the fit of the $Q$ and $U$ Stokes parameters as a function of energy with the two {\tt polconst} components for both the observations.
The contribution to the total flux of the polarized component is different (higher in Obs.~1) probably due to a dilution effect by unpolarized radiation connected with a low covering fraction of the Comptonization component (see Table~\ref{tab:spectrum}). Unfortunately, even if the source was active in the radio band during the observation campaign, it is impossible to compare the direction of the PA with the direction of the jet because the radio observations reported in this paper do not have the spatial resolution to resolve it. Moreover, the literature does not report any information about jet direction.

The variation in PA of the total emission, reported in Sect.~\ref{sec:modelindep}, obtained with the \textsc{ixpeobssim} model independent analysis can be explained by the different PAs of the disk and the Comptonized spectral components that are neither aligned nor 90\degr\ apart. The PA of the total emission varies from being positive at lower energies  (dominated by the {\tt diskbb} component) to negative values at higher energies dominated by the Comptonization component with a positive PA.

When assessing polarization as a function of time, no significant variations are seen. This implies that polarization is not sensitive to the flux variation present in the light curves (see Fig.~\ref{fig:lc}).

\subsection{Spectral energy distribution}\label{Sec:SED}

\begin{figure}
\centering
\includegraphics[scale=.23]{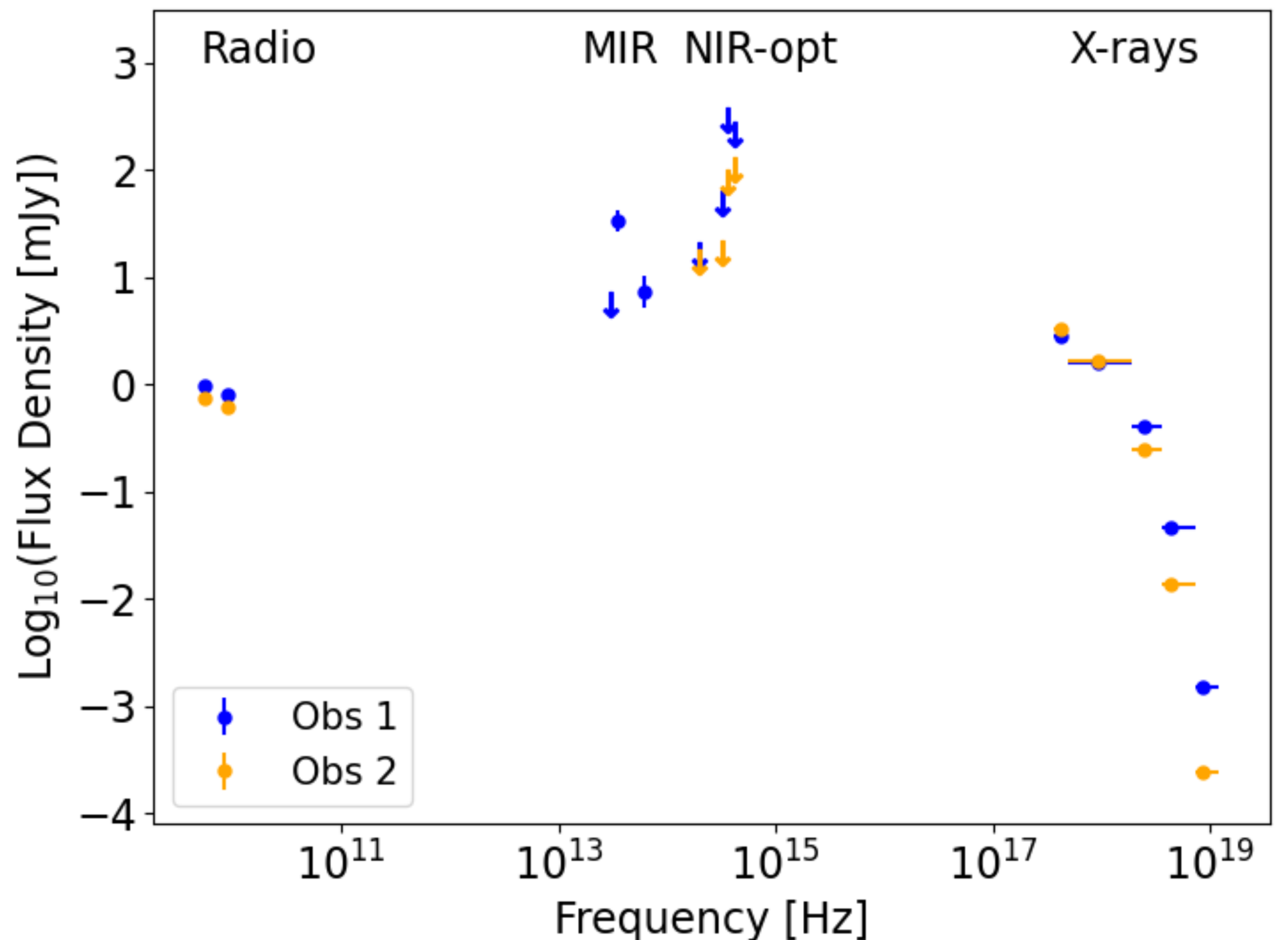}
\caption{Broadband radio to X-ray SED of \gx during the two  observations. For Obs.~2 no mid-IR flux is reported as observations were not performed with VISIR at that time (see Sect. \ref{sec:visir} for details).
The flux densities at mid-IR, near-IR, and optical frequencies have been de-reddened (as described in Sect. \ref{Sec:SED}). The errors are at 68\% confidence level.}
\label{fig:SED}
\end{figure}

Thanks to the multi-energy observation campaign it was possible to produce a broadband (radio--X-ray) spectral energy distribution (SED) of \gx for both the observations (see Fig.~\ref{fig:SED}).
Optical and infrared fluxes were de-reddened using the hydrogen column density reported in this work ($N_{\rm H}=4.93\times10^{22}$\,cm$^{-2}$ and $4.88\times10^{22}\, \rm cm^{-2}$ in Obs.~1 and 2, respectively), which was converted into an estimate of the $V$-band extinction $A_V$ using the relation reported in \citet{Foight2016},
resulting in $A_{V}=17.18\pm0.77$ mag and $17.00\pm0.72$ mag for the two observations, respectively. The different absorption coefficients at optical and near-IR wavelengths were then   evaluated using the relations reported in \citet{Cardelli1989} and \citet{Nishiyama_08}, respectively. For the mid-IR, the coefficients reported in \citet{Weingartner2001} were used instead.

\gx is detected at radio and mid-IR wavelengths, with upper limits in the near-IR and optical. As mentioned in Sect.~\ref{sec:rem}, this is due to the high value of dust extinction of $A_{V}\approx17$. This implies an extinction of $\sim$7 mag at 1\,$\mu$m ($Y$-band) and $\sim$3 mag at 1.6\,$\mu$m ($H$-band). The radio spectral index of $-0.4 \pm 0.1$ is too steep for the radio emission to arise from a steady compact jet (for which we expect a spectral index from $\sim 0$ to +0.5). As such, it is likely due to optically thin jet ejections or to a combination of compact jet and optically thin ejections. In the mid-IR we report the first detection of this source. The de-reddened flux densities are comparable to the near-IR values reported in the literature \citep{Naylor1991,Jonker2000,Bandyopadhyay2003}; however, the de-reddened near-IR fluxes depend sensitively on the value of the interstellar extinction. 
We note that different values of the neutral hydrogen column densities are reported in the literature, ranging from $N_{\rm H} = 2.54 \times 10^{22}$ cm$^{-2}$ to $6.20 \times 10^{22}$ cm$^{-2}$ \citep{Christian1997,Zeegers2017,Homan2018,Clark2018,Bhulla2019}. 
\citet{Yang2022} has derived $N_{\rm H}=(4.52 \pm 0.01)\times 10^{22}$ cm$^{-2}$ by measuring the Si~K edge due to scattering by dust using the \textit{Chandra} gratings. 
This value is in agreement with our measurements because $N_{\rm H}$ is slightly overestimated when fitting the continuum with the absorption model \texttt{tbabs} \citep{Corrales2016}. 
In this work we observed \gx over a significantly larger energy range, and we are confident to have properly constrained the absorption in the interstellar medium. However, if there is a component of the neutral hydrogen column that is intrinsic to the LMXB, this could cause varying measurements and introduce uncertainty into the relation between $N_{\rm H}$ and the extinction $A_{V}$.
Moreover, older works using \citet{Anders89} solar abundances rather than \citet{Wilms2000} interstellar abundances, could be affected by model systematics.

The mid-IR flux measured with the VLT VISIR is higher than the extrapolation of the ATCA spectral index from radio to mid-IR. The mid-IR emission is therefore unlikely to be due to optically thin synchrotron emission from discrete jet ejections that were seen in the radio, but it could be optically thin synchrotron emission from the compact jet (from above the jet spectral break; \citealt{2013MNRAS.429..815R}). We find evidence of strong mid-IR variability between the two epochs, with the 9--11\,$\mu$m flux density changing from $< 3.27$ mJy on March 28, 2023, to $9.95 \pm 2.31$\,mJy on March 31, 2023. While this variation by a factor of $\ge 3$ ($\ge 2.5$ magnitudes) is quite high for a LMXB accretion disk on these timescales, high amplitude (spanning several magnitudes) infrared variability has been reported from a number of bright persistent NS-LMXBs, including GX~17+2, Cir~X-1, 4U~1705$-$440, and GX~13+1 \citep[e.g.,][]{Glass1994,Callanan2002,Bandyopadhyay2002,Homan2009,Corbet2010,Harrison2011}. This variability has   generally been interpreted as indicative of highly variable synchrotron emission from a compact jet. Variable near-IR polarization has also been reported from the bright persistent NS-LMXBs Sco X-1 and Cyg~X-2, and from the NS-LMXB transient SAX J1808.4$-$3658 \citep{Shahbaz2008,Russell2008,Baglio2020}. The de-reddened mid-IR 4.7--8.7 $\mu$m spectral index on March 03, 2023, is $-2.3$ to $-0.4$ (adopting $A_{V}=17.18\pm0.77$), which is consistent with optically thin synchrotron emission from relativistic particles or with a steeper particle distribution with a thermal (Maxwellian) component, as has been seen at infrared wavelengths from jets in some black hole LMXBs \citep[e.g.,][]{Russell2010,Shahbaz2013}. Thus, this emission could originate from a compact jet that peaks in the mid- or far-IR, although follow-up observations characterizing the IR spectrum and variability would be beneficial to confirm the nature of the IR emission. If the compact jet is present, its spectrum from radio to mid-IR must be inverted, with index $> 0.33$ (and fainter than the observed radio emission). The radio to IR spectrum is similar in some ways to the NS-LMXBs 4U 1728$-$34 and 4U 0614+091 \citep{Migliari2010,DiazTrigo2017}.

\begin{figure}
\centering
\includegraphics[scale=.75]{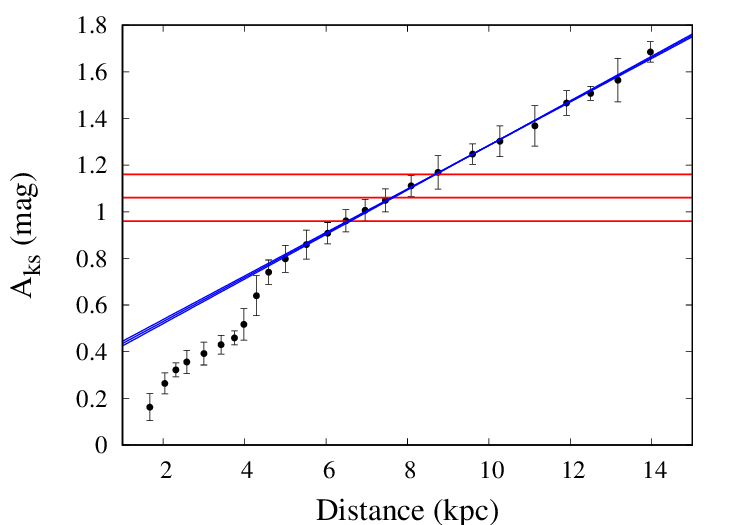}
\caption{Fit of the  infrared extinction $A_{K_s}$ between 5 and 15 kpc from \citet{Marshall_06} with a linear function  (the best-fit line and  the lines taking into account the associated errors are in blue). 
The red horizontal lines represent the estimate with the corresponding upper and lower limits of $A_{K_s}$ from Eq.~\eqref{eq:Aks}.
The distance to the source is thus $d=7.6\pm1.1$ kpc.}
    \label{fig:extinction}
\end{figure}

\subsection{Measurement of the distance to the source}\label{sec:distance}

In order to derive some spectral model parameters (namely $R_{\rm in}\sqrt{\cos\theta}$ of {\tt diskbb} and $R_{\rm bb}$ of {\tt bbodyrad}), an estimate of   the distance to the source is needed. This estimate can be obtained  from the equivalent hydrogen column density $N_{\rm H}$, which we obtain from our analysis.
We find that $N_{\rm H}= (4.93^{+0.12}_{-0.06}) \times 10^{22}$ cm$^{-2}$ and $(4.88^{+0.03}_{-0.04}) \times 10^{22}$ cm$^{-2}$ for the first and the second observations, respectively. 
Because the two values are the same at the 90\% confidence levels, we use the average value and its largest uncertainty range,   $(4.91^{+0.14}_{-0.07}) \times 10^{22}$ cm$^{-2}$,  in the following discussion. 
To estimate the distance to the source we adopt the approach proposed by \citet{Gambino2016}.
We use  the model of the infrared Galactic interstellar extinction discussed by \citet{Marshall_06}. Because the Galactic coordinates of \gx are $l=5\fdg08$ and $b=-1\fdg02$, we adopt the map that relates the infrared extinction $A_{K_s}$ with the source distance  $d$ valid for $l=5\degr$ and $b=-1\degr$ (see black dots with the error bars in Fig.~\ref{fig:extinction}).

Because the visual extinction $A_V$ is related to  $N_{\rm H}$ as \citep{Foight2016}
 \begin{equation}
     N_{\rm H}=(2.87\pm 0.12) \times 10^{21} \; A_V ,
 \end{equation}
and the relation between $A_V$ and the extinction in the $K_s$ band is \citep{Nishiyama_08} 
 \begin{equation}
     A_{K_s}=(0.062\pm 0.005)  \; A_V ,
 \end{equation}
we obtain
\begin{equation} \label{eq:Aks}
A_{K_s}=\frac{(0.062\pm 0.005)}{ (2.87\pm 0.12) \times 10^{21} }  \; N_{\rm H} \; {\rm mag} =  1.06 \pm 0.10\ \mbox{mag}. 
 \end{equation}
We fit the $A_{K_s}$ values between 5 and 15 kpc with a linear function  (the best-fit line and the lines taking into account the associated errors are in blue  in Fig. \ref{fig:extinction}). We infer that the distance to the source is $d=7.6\pm1.1$ kpc. This distance is in agreement with the previously reported values \citep{Penninx1989, Smith2006}.

\section{Discussion}
\label{sec:disc}

As shown in Sect.~\ref{sec:results}, the two multiwavelength observations of \gx allowed us to catch the source when it was covering the complete Z-track on its CCD--HID (see Fig.~\ref{fig:ccdnustar}). In particular, during the first observation \gx was on the HB of the track, while in the second observation it moved across to the NB and FB. Very interestingly, we found the same behavior in the peculiar transient XTE~J1701$-$462 \citep{Cocchi2023}, where the PD was correlated with the source position on the Z-track, being higher in the HB and decreasing by a factor of about two in the NB. \citet{Long22} hypothesized the same behavior also for Sco~X-1.
There are at least three regions that may potentially contribute to the polarization: the BL--SL, the accretion disk, and the
reflection of the BL--SL photons from the disk atmosphere or a wind.

\begin{table} 
\centering
\caption{Best-fit parameters with reflection included in the energy spectrum model of \gx from \NICER, \Nustar, and \INTEGRAL data simultaneous to \IXPE observations.} 
\footnotesize
\begin{tabular}{ll c c  }
\hline
\hline
Components & Parameters & Obs. 1 & Obs. 2   \\
\hline 
{\tt edge} & $E$ (keV) & $1.820\pm0.013$ & $1.818\pm0.013$ \\
& $\tau$ & $0.15\pm0.02$ &$0.15\pm0.02$\\ 
{\tt edge} & $E$ (keV) & \multicolumn{2}{c}{1.952 (fixed)}\\
& $\tau$ & $0.05\pm0.02$ &$0.05\pm0.02$ \\     
{\tt edge} & $E$ (keV) & \multicolumn{2}{c}{2.280 (fixed)}\\
 & $\tau$ & $0.034\pm0.014$ & $0.027\pm0.013$ \\  
{\tt edge} & $E$ (keV) & 
\multicolumn{2}{c}{2.444   (fixed)}\\ 
& $\tau$ &$0.049\pm0.013$ & $0.038\pm0.013$ \\ 
{\tt edge} & $E$ (keV) & \multicolumn{2}{c}{3.161   (fixed)}\\  
& $\tau$ & $0.021\pm0.007$ & $0.011\pm0.007$ \\    
{\tt tbabs}  & $N_{\rm H}$ ($10^{22}$  cm$^{-2}$)& $4.80\pm 0.05   $ & $4.87 \pm 0.05  $   \\
{\tt diskbb} &  $kT$  (keV) & $1.21\pm 0.05$ & $1.20 \pm 0.03$ \\
&  $R_{\rm in}\sqrt{\cos{\theta}}$  (km)\tablefootmark{a} & $16\pm1$ & $19\pm1$ \\
{\tt comptb} & $kT_0$ (keV) &
$1.93\pm 0.13   $ & $1.71 \pm 0.04  $   \\ 
& $\alpha$  & $2.9^{+1.7}_{-0.8}$ & $<1.16$  \\
 & $kT_{\rm e}$ (keV)  & $3.6^{+1.9}_{-0.4} $ &  $3.24^{+0.38}_{-0.10}$ \\
 &  $\log A$   & 8 (fixed)&
$-1.54 ^{+0.54}_{-0.06}$  \\
& $N_{\rm comptb}$   & $0.148 \pm 0.003 $ &  $0.123 \pm 0.006$\\
{\tt expabs}   & $E_{\rm cut}$ (keV) & [= $kT_{0}$] & --\\
{\tt powerlaw}   & $\Gamma $ &  $2.4$ (fixed)  &  --  \\
& norm & $0.17\pm 0.05$  & -- \\      
{\tt rdblur} & betor  &
\multicolumn{2}{c}{ $-2.5$ (fixed)}   \\ 
 & $R_{\rm in}$ ($R_{\rm g}$)  & \multicolumn{2}{c}{ 7 (fixed)}   \\
& $R_{\rm out}$ ($R_{\rm g}$) & \multicolumn{2}{c}{ 1000 (fixed)}    \\
& incl. (deg) & \multicolumn{2}{c}{  $60$    (fixed)}    \\
{\tt rfxconv} & rel-refl  &\multicolumn{2}{c}{0.3 (fixed)}   \\ 
& $\log \xi$  & \multicolumn{2}{c}{ 3.8 (fixed)}   \\
&  $\chi^2$/d.o.f.  &  268/295 &  247/260   \\ 
\hline
\end{tabular}
\tablefoot{The errors are at 90\% confidence level. \tablefoottext{a}{Radii are estimated assuming the distance to the source of 7.6~kpc (see Sect.~\ref{sec:distance}).}  }     
\label{tab:reflectionspectrum}
\end{table}

The spectropolarimetric analysis of the \IXPE data (Table \ref{tab:gainfitixpe}) shows that the disk polarization is about 2\% in both observations, which is compatible with the classical results of a high optical depth scattering atmosphere at an inclination of 60\degr\ \citep{chandrasekhar1960}.
The higher PD value of the hard component, on the other hand, cannot be explained by repeated Compton scattering
in high optical depth environments, neither for a boundary layer coplanar with the disk (which otherwise
would resemble a Chandrasekhar-like slab) nor for a spreading layer around the neutron star, for which
a maximum PD of $\la 2$\% is expected.

Disk reflection is probably the most natural way to explain a PD on the  order of 4\%--5\%, as shown by \citet{LS85}. 
It is important to note that we do not find a strong reflection signature in the spectral analysis, and it deserves mentioning that the reflection contribution to the spectrum may be low, and sometimes may be embedded in the continuum, but can nevertheless make a large  contribution to the net polarization signal \citep{schnittman&krolik2009}. This may be particularly true if the primary spectrum is not a hard power law, but a blackbody-like spectrum with a rollover below 30 keV. We note that a similar argument has been used also for Cyg~X-2 \citep{Farinelli23}  and  GX~9+9 \citep{Ursini2023} to explain the high PD attributed to the Comptonization spectrum in a twofold spectropolarimetric approach.

The PD of the reflected radiation is not easy to predict because it depends on geometrical and physical factors (e.g., the disk ionization parameter); however, it is not likely to exceed $\sim$20\% \citep{matt1993,poutanen1996,schnittman&krolik2009}.
We tested the hypothesis of whether there is a reflection component in the \gx energy spectrum by making some assumptions: (1) a highly ionized disk due to the absence of a broad  emission line in the Fe-K region of the spectrum; (2) inclination $i=60\degr$ since \gx is a Cyg-like Z-source \citep{Homan2018}; (3)   the reflection amplitude $f=\Omega/2\pi$  kept at a typical value for the NS-LMXBs  \citep[namely 30\%; see, e.g.,][]{Disalvo_15,Matranga_17}. 
We applied the following spectral models:
\begin{description}
\item[Obs.~1:] {\tt constant*tbabs*(diskbb + expabs*powerlaw + rdblur*rfxconv*comptb + comptb)},
\item[Obs.~2:] {\tt constant*tbabs*(diskbb + rdblur*rfxconv*comptb + comptb)}.
\end{description}
The {\tt comptb} model~\citep{Farinelli2008} component was included in place of the {\tt thcomp} to prevent a double convolution of the BL--SL blackbody. Details on {\tt  rfxconv} and {\tt rdblur} model components can be found in~\citet{Kolehmainen2011} and ~\citet{Fabian1989}, respectively.
The sum {\tt rdblur*rfxconv*comptb + comptb} accounts for the reflected radiation (the first term) plus the incident radiation (the second term). 

This modeling of the reflection component contribute $\sim$22\% of the flux in Obs.~1 and $\sim$12\% in Obs.~2. This contribution to the total emission can easily account for the polarization detected (see Table~\ref{tab:reflectionspectrum} for details of the fit parameters of the energy spectrum).
The fraction of Comptonized photons in Obs.~2 is significantly smaller with respect to Obs.~1, as also obtained  in Sect.~\ref{sec:results} by applying only the Comptonization model {\tt thcomp}. The  parameter $\log A=-1.54^{+0.54}_{-0.06}$  in {\tt comptb}  corresponds to a $2.8^{+6.3 }_{-0.4}\%$ of photons upscattered in energy (consistent with $3.2^{+8.0}_{-0.5}\%$ obtained when reflection is not taken into account as in Sect.~\ref{sec:results}).
This confirms the presence of a blackbody component observed through an almost vanished Comptonization medium in Obs.~2, even if reflection is included. However, the fraction of Comptonized photons, albeit small ($\sim$3\%), is still sufficient to manifest its presence at high energy. If we  were to neglect this small fraction of Comptonized photons by substituting {\tt comptb} with a blackbody component, such as {\tt bbodyrad}, we would obtain an unacceptable fit result ($\chi^2_{\nu}=3.6$, with a significant excess at high energy due to this small fraction of Comptonized photons).

Another possible mechanism for producing polarization is related to scattering in the wind above the accretion disk. As was shown in \citet{ST85}, the emission scattered once in a plane (e.g., equatorial wind) can be polarized up to 27\% for an inclination $i=60\degr$. 
This polarization degree is only weakly dependent on the opening angle of the wind.
Assuming that 20\% of the source emission is scattered, we can obtain the observed PD.
Recently, a similar model was shown to explain well the presence of a constant polarized component in the X-ray pulsars RX~J0440.9+4431 / LS~V~+44~17 \citep{Doroshenko2023}. 
Although it is well known that strong winds can   indeed be present in the soft states of X-ray binaries \citep[e.g.,][]{NeilsenLee2009, Ponti2012, Ponti2014}, it is worth noting that we do not see wind absorption features in the energy spectrum of \gx, implying that, if present, the wind should be completely ionized.  

Both \IXPE observations show similar behavior of the PD values: they are smaller in the 2--4~keV band and increase slightly with energy. The reduction of the PD at lower energies can be explained by the energy dependence of the disk emission. In the low-energy band we have a marginally polarized disk emission dominating the spectrum, while at higher energies the emission of the SL and/or the scattered or reflected component is more visible.
Another feature of the observed emission that needs to be addressed is the variation in PA with energy. 
In the spectropolarimetric analysis the disk and the Comptonization components have nonorthogonal PAs, and thus the variation in polarization plane of the total emission with energy can be interpreted as the energy-dependent contribution of these two components to the total emission. 
The same applies if, for instance, the SL emission is scattered in the wind and the disk emission is polarized with a different and nonorthogonal angle. The PA of the combined emission will be energy dependent.
It is well known that the disk emission exhibits a rotation of the polarization plane with energy \citep{Connors1980,Dovciak2008,LVP22}.
Simulations of the SL emission also show a change in PA with energy, but predict a PD at most of 1.5\% (Bobrikova A., in prep.).
Thus, a single component explanation of the PA variation cannot be considered compatible with the high measured PD.
Further modeling will be needed to satisfactorily address the explanations presented above.

\section{Summary}
\label{sec:summary}

\IXPE observed the NS-LMXB \gx  twice in the period March--April 2023. 
Contemporary observations in the X-ray energy band were put in place with \NICER, \Nustar, and \INTEGRAL. 
Multiwavelength coverage was ensured by \ATCA in the radio, VLT VISIR in mid-IR,  REM in optical, and NIR and LCO in the optical. 

During the observations \gx moved across the entire Z-track. 
\Nustar clearly disentangled  the NB with respect to the FB, thanks to its extended energy band. 
The presence of a hard tail,  reported in previous analyses, was clearly detected in Obs.~1, but not in Obs.~2, when the source had a softer energy spectrum. 
The X-ray PD was $\sim$4\% during Obs.~1 when the source  was in the HB and $\sim$2\% during Obs.~2  with the source in the NB-FB. 
This result is in agreement with findings from the other Z-sources  observed by \IXPE (namely Cyg~X-2 and XTE~J1701$-$462). 

The source manifested an unexpected variation in PA as a function of energy, by $\sim$ 20\degr. 
The magnitude of the variation combined with the magnitude of the PD require further modeling. However, it is likely related to the different PAs of the disk and Comptonization components, which are nonorthogonal and, moreover,  have  emission peaks at different energies.

In the radio band the source was detected, but only upper limits to the polarization were obtained ($\sim$6\% at 5.5~GHz and 9~GHz in Obs.~1 and 12.5\% at 5.5~GHz and 20\% at 9~GHz in Obs.~2). The mid-IR counterpart was detected in M  and in the J8.9 bands. This emission could originate from a compact jet that peaks in the mid- or far-IR.
Follow-up observations characterizing the IR spectrum and its variability would be beneficial.
Due to the very high extinction toward the source and to the crowded field, the source was not detected in any of the optical and NIR images acquired.

\begin{acknowledgements}
The Imaging X-ray Polarimetry Explorer (IXPE) is a joint US and Italian mission.  The US contribution is supported by the National Aeronautics and Space Administration (NASA) and led and managed by its Marshall Space Flight Center (MSFC), with industry partner Ball Aerospace (contract NNM15AA18C).  The Italian contribution is supported by the Italian Space Agency (Agenzia Spaziale Italiana, ASI) through contract ASI-OHBI-2022-13-I.0, agreements ASI-INAF-2022-19-HH.0 and ASI-INFN-2017.13-H0, and its Space Science Data Center (SSDC) with agreements ASI-INAF-2022-14-HH.0 and ASI-INFN 2021-43-HH.0, and by the Istituto Nazionale di Astrofisica (INAF) and the Istituto Nazionale di Fisica Nucleare (INFN) in Italy.  This research used data products provided by the IXPE Team (MSFC, SSDC, INAF, and INFN) and distributed with additional software tools by the High-Energy Astrophysics Science Archive Research Center (HEASARC), at NASA Goddard Space Flight Center (GSFC).

ATCA is part of the Australia Telescope National Facility (https://ror.org/05qajvd42) which is funded by the Australian Government for operation as a National Facility managed by CSIRO. We acknowledge the Gomeroi people as the Traditional Owners of the ATCA Observatory site.
Partly based on observations with INTEGRAL, an ESA project with instruments and science data centre funded by ESA member states (especially the PI countries: Denmark, France, Germany, Italy, Switzerland, Spain), and with the participation of the Russian Federation and the USA. Data have been reduced using the \href{https://www.astro.unige.ch/mmoda/}{Multi-Messenger Online Data Analysis} platform provided by the ISDC (University of Geneva), EPFL, APC, and KAU. 
J.P. acknowledges support from the Academy of Finland grant 333112. A.B. is supported by the Finnish Cultural Foundation grant 00220175.
T.D.R. thanks Jamie Stevens and James Miller-Jones for the helpful discussions regarding the radio observations.
J.S. and M.D. acknowledge the support from the GACR project 21-06825X.

This research was also supported by the INAF grant 1.05.23.05.06: ``Spin and Geometry in accreting X-ray binaries: The first mti frequency spectro-polarimetric campaign".
\end{acknowledgements}

\bibliographystyle{aa}
\bibliography{ref}

\end{document}